\documentclass[a4paper,fleqn,usenatbib,useAMS]{mnras}

\usepackage{newtxtext,newtxmath}

\usepackage[T1]{fontenc}
\usepackage{ae,aecompl}

\usepackage{graphicx}   
\usepackage{verbatim}
\usepackage{color}
\usepackage{lineno}
\usepackage{cleveref}
\usepackage{lineno}

\newcommand{\cc}{\mathrm{c}}
\newcommand{\dd}{\mathrm{d}}

\newcommand{\gauss}[3]{\frac{1}{\sqrt{#3}}\exp \left(-\frac12
    \frac{\left( #1 - #2 \right)^2}{#3} \right)}

\title[CF3 velocity field]{The peculiar velocity field up to
  $z \sim 0.05$ by forward-modeling 
  \emph{Cosmicflows-3} data}

\author[R. Graziani et al.]{
R. Graziani$^{1,2}$,
H. M. Courtois$^{1}$,
G. Lavaux$^{3}$,
Y. Hoffman$^{4}$,
R. B. Tully$^{5}$,
\newauthor
Y. Copin$^{1}$,
D. Pomar\`ede$^{6}$
\\
$^{1}$ University of Lyon, UCB Lyon 1, CNRS/IN2P3, IPN Lyon, France \\
$^{2}$ Universit\'e Clermont Auvergne, CNRS/IN2P3, Laboratoire de
Physique de Clermont, France \\
$^{3}$ Institut d'Astrophysique de Paris, Sorbonne Université/CNRS,
France \\
$^{4}$ Racah Institute of Physics, Hebrew University, Jerusalem 91904, Israel\\
$^{5}$ Institute for Astronomy, University of Hawaii, 2680 Woodlawn
Drive, Honolulu, HI 96822, USA \\
$^{6}$ Institut de Recherche sur les Lois Fondamentales de l'Univers, CEA Universit\'e Paris-Saclay, France
}

\date{Accepted XXX. Received YYY; in original form ZZZ}

\pubyear{2018}

\begin{document}

\label{firstpage}
\pagerange{\pageref{firstpage}--\pageref{lastpage}}
\maketitle

\begin{abstract}

A hierarchical Bayesian model is applied to the \emph{Cosmicflows}-3
catalog of galaxy distances in order to derive 
the peculiar velocity field and distribution of matter within $z \sim
0.054$. The model assumes the $\Lambda$CDM model within the linear
regime and includes
the fit of the galaxy distances together with the underlying density
field. By forward modeling the data, the method is
able to mitigate biases inherent to peculiar velocity analyses,
such as the Homogeneous Malmquist bias or the log-normal distribution of
peculiar velocities. The statistical uncertainty on the
recovered velocity field is about 150 km/s depending on the location,
and we study systematics coming from the selection function and
calibration of distance indicators. The resulting velocity field and related density fields recover the cosmography of the
Local Universe which is  presented in an unprecedented volume of universe
10 times larger than previously reached.
This methodology open the doors to reconstruction of initial conditions for larger and more accurate constrained cosmological
simulations. This work is also preparatory to larger peculiar velocity
datasets coming from
Wallaby, TAIPAN or LSST.
\end{abstract}

\begin{keywords}
large-scale structure of Universe, dark matter, observations, galaxies:
distances and redshifts, methods: data analysis
\end{keywords}



\section{Introduction}
\label{sec:introduction}
Peculiar motions of
galaxies are due to the gravitational interaction with the underlying
density field of matter. Thus peculiar velocities of galaxies are a
powerful and unbiased tool to study the dynamics and structure of the Local Universe. 
Velocities have been used as probes of cosmological parameters (e.g.~\citet{1997ApJ...486...21Z,2011ApJ...736...93N,2017MNRAS.468.1420F,2017MNRAS.471.3135H,2017MNRAS.470..445N,2018MNRAS.480.5332W}), for
cosmography studies~\citep{1999ApJ...522....1D,Tully:2014gfa,2017arXiv170807547C,2017NatAs...1E..36H}, and to
set initial conditions for constrained simulations~\citep{2010arXiv1005.2687G,2016MNRAS.455.2078S}. 
In the past, the Local Universe peculiar
velocity field has been reconstructed from the expected response to the observed redshift distribution of galaxies taken 
as tracers of the mass distribution~\citep{2011ApJ...736...93N,2014MNRAS.442.1117D,2004MNRAS.352...61H,2014MNRAS.445..402H,2016MNRAS.455..386S}. \\

An alternate approach has been followed by the Cosmicflows program.
Peculiar velocities are inferred from departures of measured distances from the expectations of uniform cosmic expansion.  
Cosmicflows catalogs~\citep{2008ApJ...676..184T,2013AJ....146...86T} have been analyzed
through the Wiener Filter/Constrained
Realizations methodology (WF/CR)~\citep{1999ApJ...520..413Z,2012ApJ...744...43C}. 
The assumption is made with the WF/CR methodology that
the measured velocities have a Gaussian noise and that their
2-point correlations are given by the $\Lambda$CDM model. \\
 
It is critical that steps be taken to mitigate the Malmquist bias that arises 
from errors in distance.  Objects are preferentially misplaced from regions of 
higher sampling density to lower sampling density~\citep{1995PhR...261..271S}.
As an example, it can be anticipated that galaxies assigned the greatest distances
probably have large positive errors.  These galaxies will be attributed with
large negative peculiar velocities. The main objective of the present
paper is to explore a more rigorous solution than the previous approach
of~\citet{2015MNRAS.449.4494H} in order to overcome this issue.

As a framework,~\citet{2016MNRAS.457..172L} has developed a fully Bayesian algorithm that
incorporates the constrained realizations technique within a 
statistical model accounting for the uncertainty on the location of
tracers. We use here a similar method to reconstruct the 3D linear
velocity field from Cosmicflows-3 data up to redshift $z\sim 0.054$. \\

The paper is organized as follows: the
Cosmicflows-3 data is briefly described in Section~\ref{sec:data} then
the method is detailed in Section~\ref{sec:method}. The results includes
an analysis on the reconstruction of the linear velocity field in
Section~\ref{sec:results} and an overlook on the resulting cosmography
is given in Section~\ref{sec:cosmography}. There are comments on
outstanding issues in Section~\ref{sec:discussion}.

\section{Data}\label{sec:data}
\subsection{Compilation of distance moduli}
The reconstruction is based on the Cosmicflows-3 (CF3) catalog~\citep{2016AJ....152...50T}\footnote{Available as a table at http://edd.ifa.hawaii.edu}. CF3 provides a compilation of almost 18,000 galaxy distances,
with the high redshift ones computed for the most part from three methodologies: the luminosity-linewidth
relation of spiral galaxies, TF \citep{1977A&A....54..661T}, the Fundamental Plane of early-type
galaxies \citep{1987ApJ...313...59D,1987ApJ...313...42D}, and Type Ia
supernovae \citep{1993ApJ...413L.105P}. The absolute scale of the global distance ladder
of these methodologies is given by galaxies that overlap with
Cepheid variables \citep{1912HarCi.173....1L} or tip of the Red Giant
Branch \citep{1990AJ....100..162D}. The Surface Brightness
Fluctuation \citep{1988AJ.....96..807T}  method helps providing a bridge between the near and far
field. 
The CF3 compilation is heterogeneous, unlike concurrent single methodology samples \citep{2007ApJS..172..599S, 2014MNRAS.445.2677S,  2014MNRAS.445..402H}.  
Indeed, when available, CF3 incorporates the major literature contributions.  
For inclusion in CF3, a source must usefully complement other components while overlapping 
sufficiently to assure consistency of scale.  Each linkage has associated uncertainties.  
However, what is lost in the ambiguities of linkages is surely more than compensated by the 
dynamic range of the CF3 catalog.  Nearby, coverage is dense and distances are accurate at the level of 5\%.  
Farther away, Fundamental Plane contributions emphasize coverage of major clusters.  
TF samples preferentially provide distances to galaxies in the field.  SNIa hosts are scattered serendipitously.  
The methodologies converge in groups where there can be multiple contributions.  
A group, or an individual galaxy where there is a convergence, has a unique distance.  
Averaged over all such cases, methodologies should agree.

\subsection{Groups}\label{sec:groups}
Our goal is to derive the linear velocity field from the peculiar velocities
of galaxies. However, galaxies in groups or
clusters are affected by non linear motions which are not
 modeled within our $\Lambda$CDM linear framework. 
Our solution is to average information over the small scale of groups.
\citet{2015AJ....149..171T} provides a catalog of groups built from 
the 2MASS redshift survey complete to $K_s= 11.75$ \citep{2012ApJS..199...26H}.
Candidate galaxies are either directly linked to these groups as members of the 2MASS sample
or indirectly linked by close spatial and velocity association.
A group is assigned a velocity that averages over all known members and
a distance that is the weighted average over those constituents with the necessary measurements.
Uncertainties with $N$ measures are reduced roughly as $\sqrt{N}$ (depending on the details of 
the contributing methodologies), so groups are particularly high value entries in the CF3
catalog.

\subsection{Selection function}\label{sec:selec}
\begin{figure}
  \centering
   \includegraphics[width=1\linewidth]{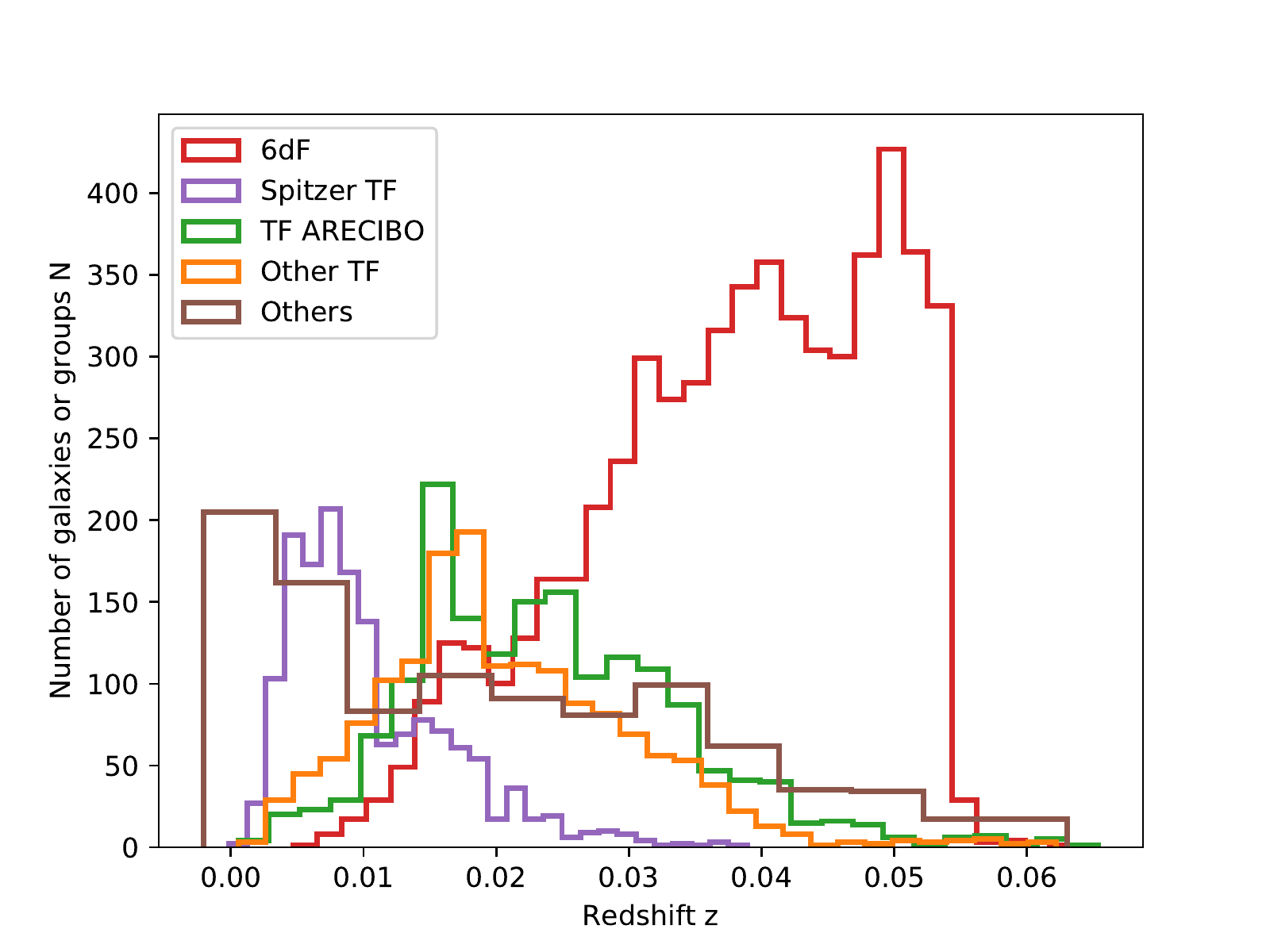}

  \caption{Histogram of the observed redshifts for
  the five subsamples described in Section~\ref{sec:groups}.}
  \label{fig:z}
\end{figure}

Our Bayesian methodology (see Section~\ref{sec:method}) needs priors on the 
statistical distribution of distances. Because it is a composite catalog, CF3 is
inhomogeneous both in distance and angular coverage. Consequently it does
not admit a unique and simple selection
function.
Still, it is possible to identify 5 main subsamples based on the
original observational surveys:
\begin{enumerate}
\item 6dFGSv data provides the most well defined subsample: it has a high
  degree of completeness up to a sharp cutoff at $z = 0.054$. The subsample contains 5,777 galaxies or groups with a median redshift
  of $z =0.039$.
\item Another reasonably well defined subsample is based on the TF method
  with near-infrared photometry from Spitzer Space Telescope. This constituent
  gives particular emphasis to coverage at low galactic latitudes.
  The 1,546 galaxies or
  groups included in this subsample have a median redshift of $z = 0.009$.
\item TF data is particularly deep within the region covered by Arecibo Telescope which observes
  galaxies of declination $\delta \in [0,38]^{\circ}$. The subsample contains 1,628 galaxies or groups with a median redshift
  of $z =0.023$.
\item Other TF data than in the Arecibo declination range. The subsample contains 1,569 galaxies or groups with a median redshift
  of $z =0.019$.
\item Other data coming from heterogeneous
  methodologies. Cepheids, tip of the red giant branch and surface brightness fluctuation
  contributions are local and of high accuracy.  Groups which have more
  than ten members with known redshifts are included in this
  subsample. 
  Group distances are mostly averaged over Fundamental Plane and TF measures 
  with occasional SNIa contributions.  Isolated SNIa lie over a wide range of redshifts up to $z=0.1$.
  This last subsample 
  does not have a simple selection
  function. This subsample contains 963 galaxies or groups with a median redshift
  of $z =0.015$.
\end{enumerate}
The total number of tracers in the catalog is $N=11,483$. Figure~\ref{fig:z} shows the redshift histograms of all these
subsamples. We see that 6dF data plays a major role in CF3 and has a
singular behavior with a sharp cutoff at redshift $z = 0.054$. Except
for the last subsample of heterogeneous inputs, the
other subsamples redshift distributions behave as expected: a growth at
small distance due to the volume effect and then a decrease due to
selection effects. We will show in Section~\ref{sec:priors} how we model these distributions.

\section{Methodology}\label{sec:method}
This section presents the methodology developed to reconstruct the
linear velocity field underlying Cosmicflows-3 data.Our
framework is assuming a $\Lambda$CDM model and a Gaussian error
model on the observations, the distance moduli $\{\mu_i\}$ and redshifts $\{z_{i}\}$. As it will be described in the following sections, the
parameter space $\Theta$
includes the most probable luminosity distance of galaxies $\{d_{i}\}$, an effective
Hubble constant $h_{\text{eff}}$, a non-linear velocity dispersion
$\sigma_{\text{NL}}$ and the linear velocity field itself 
$\mathbfit{v}$:
\begin{equation}
  \label{eq:param_space}
  \Theta = \{\{d_{i}\},h_{\text{eff}},\sigma_{\text{NL}},\mathbfit{v}\}
\end{equation}
The general idea is to derive from basics assumptions the posterior probability of these
parameters given the data $\mathcal{P}(\Theta | \{\mu_{i},z_{i}\})$ and
then sample from it. The sampling will result in the posterior
distribution of the linear velocity field $\mathbfit{v}$ from which we
will extract the mean and standard deviation. The first
subsection reminds the reader about the $\Lambda$CDM model of peculiar
velocity statistics while the next subsections will
detail how the posterior probability is constructed and the procedure
for sampling from this complex distribution.

\subsection{Definitions and notations}\label{sec:def-not}
The methodology relies on the assumption of a flat $\Lambda$CDM model
parametrized by the set of parameters
$(\Omega_{m},H_{0})$. The $\Lambda$CDM model assumes homogeneity and
isotropy so that the cosmological redshift $\bar{z}$ of a galaxy is related to its
luminosity distance $d$ through the Hubble law:
\begin{equation}
  \label{eq:dLz}
  d(\bar{z}) = \cc \frac{1+\bar{z}}{H_{0}}\int_{0}^{\bar{z}}
  \frac{\mathrm{d} z}{\sqrt{\Omega_{\Lambda}+\Omega_{m}{(1+z)}^{3}}}
\end{equation}
where $\Omega_{\Lambda} = 1 - \Omega_{m}$. Eq.~\ref{eq:dLz} can be
numerically inverted, giving $\bar{z}(d)$. In the following we make the
dependence implicit and note $\bar{z}$ instead of $\bar{z}(d)$. The analysis assumes
the linear regime where the overdensity field $\delta(\mathbfit{r})$ is
small $|\delta(\mathbfit{r})| \ll 1$. 
 In this case the
linear theory of perturbations predicts that the
linear overdensity field is Gaussian and is described by its power spectrum $P(k)$:
\begin{equation}
  \label{eq:deltacorr}
  \langle \delta(\mathbfit{k}) \delta^{*}(\mathbfit{k}')\rangle =
   (2 \pi)^{3} \delta_{D}(\mathbfit{k} - \mathbfit{k}') P(k)
\end{equation}
where $\delta(\mathbfit{k})$ denotes the Fourier transform of
$\delta(\mathbfit{r})$ and $\delta_{D}$ is the Dirac delta
distribution. Gravitational dynamics in an expending universe dictates
that the rotational component  of the velocity decays early after the
onset of the instability and to linear order the velocity $\mathbfit{v}$
and density fields are related by:
\begin{equation}
  \label{eq:vdef}
  \nabla \cdot \mathbfit{v} = - H_{0}f\delta\left(\mathbfit{r}\right)
\end{equation}
where $f$ is the growth rate of structure and depends on the assumed
cosmological parameters. Eq.~\ref{eq:vdef} is linear as we can see
by transposing it into the Fourier space:
\begin{equation}
  \label{eq:vfourier}
  i \mathbfit{k} \cdot \mathbfit{v}(\mathbfit{k}) = - H_{0}f \delta(\mathbfit{k})
\end{equation}
Consequently, the statistic of $\mathbfit{v}$ is also Gaussian. The statistic of $\mathbfit{v}$ is described in
the Appendix~\ref{sec:line-pecul-veloc}. 

Because of the peculiar velocity field $\mathbfit{v}$, the Hubble flow
described in Eq.~\ref{eq:dLz} is distorted, and the observed redshift $z$
of a galaxy is a composition of the cosmological redshift and the
Doppler effect coming from the radial peculiar velocity:
\begin{equation}
  \label{eq:zcomp}
  1+z = \left(1+\bar{z}(d)\right)\left(1+\frac{\mathbfit{v}\left(\mathbfit{r}\right) \cdot \hat{\mathbfit{r}}}{c}\right)
\end{equation}
where $\hat{\mathbfit{r}} \triangleq \mathbfit{r}/r$. The velocity field
$\mathbfit{v}$ can be divided into a linear and a
non-linear part $\mathbfit{v}_{\text{tot}} = \mathbfit{v} +
\mathbfit{v}_{\text{NL}}$ and the non-linear part is approximated here
by an isotropic Gaussian probability distribution function:
\begin{equation}
  \label{eq:12}
  \mathcal{P}\left(\mathbfit{v}_{\text{NL}}|\sigma_{\text{NL}}\right) \sim
  \mathcal{N}(\mathbfit{v}_{\text{NL}}; \mathbfit{0},\sigma^{2}_{\text{NL}} ).
\end{equation}
where $\mathcal{N}(x;m,s^{2})$ denotes the normal distribution of mean $m$
and variance $s^{2}$ over a variable $x$ and $\mathcal{P}(x|y)$ the conditional
probability of $x$ given $y$.
From the measurements of galaxy distance moduli $\{\mu_{i}\}$ and
redshifts $\{z_{i}\}$, our goal is to infer the underlying peculiar velocity
field $\mathbfit{v}$. Since the  measurement of extragalactic distances is very noisy
(most often about 20$\%$ of relative error), we must take into account the
correlations between the observed redshifts to extract information on the
velocity field. Also, we need to take into account the possible biases
appearing when dealing with peculiar velocity field reconstruction,
which is done in Section~\ref{sec:malmquist-bias-1}. The
Section~\ref{sec:L} present how to describe the
observations given the above model.

\subsection{Malmquist bias}
\label{sec:malmquist-bias-1}
Our model is motivated by a rigourous treatment of homogeneous and
inhomogeneous Malmquist bias. The homogeneous Malmquist bias
is a
statistical bias resulting
from a combination of the volume effect nearby and the selection effects
far away. The observational uncertainties on the measured
distances scatter the observed galaxies along the radial direction. Closeby, the number of galaxies grows with the distance, so that
overall it is more likely to underestimate the galaxy distances, and as
a consequence to assign erroneously positive peculiar velocities~\citep{1995PhR...261..271S}. At large
distances, the effect is the opposite:
it is more likely that the galaxies we observe are scattered away from
us, and so in average we overestimate their distances, and assign
erroneously negative peculiar velocities. Consequently, neglecting the
homogeneous Malmquist bias would create a fake
outflow in the central region and a fake inflow on the edges. To
overcome this bias, our statistical model fits the underlying most probable
luminosity distances of galaxies, together with the velocity
field. The bias is statistically handled by attributing a prior
function to these distances. The shape of the prior function will be
detailed in Section~\ref{sec:priors}, and will be chosen so
that both the volume effect and selection effects are taken into
account.\\

The related
inhomogeneous Malmquist bias arises from structure in the observed
volume. In the
vicinity of a dense region, galaxies are more likely to be scattered
from denser to less dense regions. Consequently, the inferred
peculiar velocities are biased toward a stronger inflow onto the
structures, and thus bias the reconstructed velocity field. By
introducing the luminosity distances as parameters, our method is able to reduce this
bias. Distances are inferred by both the observations and
the reconstructed velocity field, and at each step the galaxies are
relocated with respect to the velocity field at their positions. Consequently, an inflow on a dense
region will shift the distances toward their true positions.
For a more complete analysis of the Malmquist bias, we refer the
  reader to ~\citet{1995PhR...261..271S}.

\subsection{Observed radial peculiar velocities}
\label{sec:observ-radi-pecul}
The appropriate modeling of the distribution of errors on the observed
peculiar velocities is an another important concern. The peculiar
velocities are not directly measured but are usually derived from
the distance moduli and redshifts measurements through the
Eq.~\eqref{eq:zcomp}, where the luminosity distances $d$ are computed from
the distance moduli using:
  \begin{equation}
  \label{eq:DM}
    \mu = 5 \log_{10} \frac{d}{10 ~\text{pc}}.
\end{equation}
Since the errors on the distance moduli are
supposed to be Gaussian, the resulting distribution of
peculiar velocities will not be Gaussian distributed but rather Log-normal
distributed~\citep{2016AJ....152...50T}. An example of treatment of this
effect is given by~\cite{2015MNRAS.450.1868W} who suggest the use of an unbiased
estimator of peculiar velocities with Gaussian distributed errors and which is valid at distances
$d \gtrapprox 20$ Mpc.

We use
in this paper a different approach. Instead of analyzing the observed
peculiar velocities, we rather choose to model the distance moduli observations
directly. To do so, the introduction of the luminosity distances
as free parameters allow us to take into account the Gaussian
distribution of distance moduli errors including the relativistic effects of
Eq.~\eqref{eq:zcomp}. The statistical linkage between
distance moduli and luminosity distances will be explained in Section~\ref{sec:DM}.

\subsection{Statistical model}
\label{sec:L}
We aim at recovering the linear peculiar velocity field from the CF3
observations. We work in a Bayesian framework and try to estimate the
posterior probability of $\mathbfit{v}$ given the model
described above and the observations. We proceed in three steps. First,
we detail how the likelihood of our observations $\mathcal{L}$ is
constructed (Sections~\ref{sec:DM},~\ref{sec:redshifts},~\ref{sec:likelihood}); second we impose priors on
the fitted parameters that come from the $\Lambda$CDM model presented
in the above section (Section~\ref{sec:priors}). Third, we sample from from the
posterior probability (Section~\ref{sec:sampling}).

\subsubsection{Distance moduli}
\label{sec:DM}
Distance indicators used in CF3 are expressed as distance moduli rather than
luminosity distances. The link between the two is given by Eq.~\eqref{eq:DM}.
Our primary interest in this study concerns deviations from cosmic expansion 
which are determined independently of the absolute scaling of the extragalactic distance ladder.
In addition, our analysis is insensitive to a potential monopole term that
might reflect that we live in an overall under or overdense part of the universe. 
\citet{2016AJ....152...50T} argued that the value of the Hubble Constant
consistent with the 18,000 distances in CF3 is $H_0=75$~km s$^{-1}$ Mpc$^{-1}$
to within $\sim3\%$, discounting the irrelevant absolute scaling.  
There is agreement at this level in the determination of $H_0$
between the sample within $z=0.1$ and samples of SNIa at $z\gg0.1$, limiting concern
of a substantial local monopole to flows.  

Our conclusion regarding the matter of the Hubble Constant is that the 
value is relatively well determined but uncertainties at the few percent level remain 
that are relevant for our analysis.
For this reason, we
introduce a dimensionless free parameter $h_{\text{eff}}$ to model this uncertainty in the
constant:
\begin{equation}
  \label{eq:heff}
  \mu = 5 \log_{10} \frac{d}{10 ~\text{pc}} + 5 \log_{10} h_{\text{eff}}
\end{equation}
If the absolute scale is compatible with the assumed $H_{0}$, then
$h_{\text{eff}}$ should be unity. \\

Because the  distance moduli are very noisy, we need to model
the observations by assuming that the error is Gaussian (noted
$\sigma_{\mu}$) and choose to fit for the underlying true luminosity
distance $d$ :
\begin{equation}
  \label{eq:Pmu}
  \mathcal{P}\left(\mu | d,\sigma_{\mu}\right) =
  \mathcal{N}\left(\mu; 5 \log_{10}\frac{h_{\text{eff}} d}{10 ~\text{pc}},\sigma^{2}_{\mu}\right)
\end{equation}
The parameter $h_{\text{eff}}$ is correlated to the reconstructed
velocity field and is prone to systematics coming from shifts
between zero-point scales of different methodologies, non-linear effects, selection functions and possible external bulk
flows. For this reason, one needs to take this parameter with caution
and not as direct measure of $H_{0}$. 
For the location of a galaxy in space, the angular position is measured essentially without
error. Hence, from the distance $d$ and the angular position, we can
compute the spatial position of a galaxy, which we denote $\mathbfit{r}$.

\subsubsection{Redshifts}
\label{sec:redshifts}

Eq.~\ref{eq:zcomp} gives the relation between the observed redshift
$z$ and the cosmological redshift $\bar{z}$, which can be computed from
the luminosity distance $d$ through Eq.\ref{eq:dLz}. From these
equations, we can compute the radial peculiar velocity of a galaxy:
\begin{equation}
  \label{eq:vr}
  v^{r}\left(z,d\right) = c \frac{z-\bar{z}(d)}{1+\bar{z}(d)}
\end{equation}
In CF3, the errors on individual redshifts are not provided, and we
chose to assume
that they are measured with a Gaussian error of $c \sigma_{z} =
\sigma_{cz} =
50$ km/s, the typical error for a spectroscopic redshift measurement.

To model the redshifts measurements, we introduce the
underlying linear velocity field
$\mathbfit{v} \triangleq \mathbfit{v}(\mathbfit{r}_{j})_{j \in [0,M^{3}]}$ sampled on a
grid of size $M^3$ and volume $L^{3}$. Since the sampled velocity field is
linear, we need to model the departure of the observed velocities from
the linearity. We do so by introducing a Gaussian dispersion
$\sigma_{\text{NL}}$ around the linear field which is to be evaluated.The introduction of a unique parameter
  $\sigma_{\text{NL}}$ hence models the departure of the
  overall velocity field from linearity and does not model high
  dispersion inherent to non-linear environment such as clusters of
  galaxies. For this reason the use of the grouped CF3 catalog is
  mandatory. Applying this model on the non-grouped catalog would
underestimate the redshift errors and lead to unphysical results near
high density regions.\\
 
The probability of observing the
redshift $z$ knowing the luminosity distance $d$ and the velocity
field $\mathbfit{v}$ is:
\begin{equation}
  \label{eq:Pz}
  \mathcal{P}\left(z | \mathbfit{r},d,\sigma_{z},\mathbfit{v}\right) =
  \mathcal{N}\left(v^{r}(z,d); \mathbfit{v}(\mathbfit{r}) \cdot
    \hat{\mathbfit{r}}, \sigma^{2}_{cz}(1+\bar{z})^{-2} + \sigma^{2}_{\text{NL}}\right)
\end{equation}
where $v^{r}(z,d)$ is defined by Eq.~\ref{eq:vr}. Note that $v^{r}$ is here a function of the parameter $d$ and the
observation $z$, and does not correspond to a peculiar
velocity measurement. The probability distribution above describes the
statistical link between the redshifts and the model's parameters.
In practice, the linear velocity $\mathbfit{v}$ in Eq.~\eqref{eq:Pz} is computed
on a finite-size grid $\lbrace \mathbfit{r}_{j} \rbrace$. To evaluate it at any position (such as the
position of a galaxy), we use trilinear interpolation between grid
cells. 
We note from the error distribution of Eq.~\eqref{eq:Pz} that
$\sigma_{cz}$ and $\sigma_{\text{NL}}$ are strongly correlated, the two
dispersions being different by only the relativistic factor $1+\bar{z}
\simeq 1$. As a consequence, an over (or under) estimation of $\sigma_{cz}$ will be
transferred to a under (or over) estimation of $\sigma_{\text{NL}}$, and
our results will be insensitive to the choice of $\sigma_{cz}$.

\subsubsection{Likelihood}\label{sec:likelihood}
The likelihood gives the conditional probability of our observations, namely the observed
redshifts $z$ and distance moduli $\mu$, given the model and its parameters, namely the
true luminosity distance $d$, the linear velocity field sampled on a
grid $\mathbfit{v}$, $h_{\text{eff}}$, and 
$\sigma_{\text{NL}}$. Since the conditional probabilities on $\mu$ and
$z$ are independent, the likelihood is given by the combination of
Eqs.~\eqref{eq:Pmu} and ~\eqref{eq:Pz}:
\begin{equation}\label{eq:L}
  \begin{split}
      &\mathcal{L} = \mathcal{P}\left(\{\mu_{i},z_{i}\}
        |
        \{d_{i},\sigma_{\mu,i},\sigma_{z,i}\},h_{
\text{eff}},\sigma_{\text{NL}},\mathbfit{v}\right)=\\
& \prod_{i} \mathcal{P}\left(\mu_{i} |
  d_{i},\sigma_{\mu,i},h_{\text{eff}}\right) \times \mathcal{P}\left(z_{i} | \mathbfit{r}_{i},d_{i},\sigma_{z},\mathbfit{v}\right)=\\
      &\frac{1}{2\pi}\prod_{i} \gauss{\mu_{i}}{5 \log_{10}
      \frac{h_{\text{eff}}d_{i}}{10~\text{pc}}}{\sigma_{\mu,i}^{2}}\times\\ 
      &
      \gauss{v^{r}(z_{i},d_{i})}{\mathbfit{v}(\mathbfit{r}_{i})\cdot
        \hat{\mathbfit{r}}_{i}}{\sigma_{cz}^{2}(1+\bar{z}_{i})^{-2}+ \sigma^{2}_{NL}}
  \end{split}
\end{equation}
where the index $i$ denotes the $i^{th}$ galaxy or group. 
The likelihood defined by
Eq.~\eqref{eq:L} is similar to the one in \citet{2016MNRAS.457..172L}
but simplified
because we do not aim to fit the power spectrum properties such as the
shape or the normalization. Another difference is that we directly sample
the velocity field rather than the overdensity field. We do so to avoid
the use of Fourier Transform to compute Eq.~\eqref{eq:vdef} and hence to
not be subject to periodic boundary conditions. Also, we use a trilinear interpolation to compute
the linear velocity field at any point in space, while \citet{2016MNRAS.457..172L} developed
a Fourier-Taylor algorithm to interpolate using only series of Fast
Fourier Transforms. We note that our way of proceeding does not enforce the curl-free properties of the linear
velocity field. We \emph{a posteriori} check that the curl part
of the reconstructed field is negligible.\\

From the likelihood~\eqref{eq:L}, we can estimate the probability of a given
velocity field $\mathbfit{v}$ (and associated parameters) from the
measurements of distance moduli and redshifts. This probability is given
by the Bayes theorem:
\begin{equation}
  \label{eq:bayes}
   \mathcal{P}\left(\{d_{i}\},h_{\text{eff}},\sigma_{\text{NL}},\mathbfit{v}|\{\mu_{i},z_{i}\}\right)   \propto \mathcal{L} \times \mathcal{P}(\{d_{i}\}) \mathcal{P}(h_{\text{eff}})
   \mathcal{P}(\sigma_{\text{NL}}) \mathcal{P}(\mathbfit{v})
\end{equation}
where $\mathcal{P}(\theta)$ denotes the prior on the parameter $\theta$,
$\theta \in [\{d_{i}\},h_{\text{eff}},\sigma_{\text{NL}},\mathbfit{v}]$.
The velocity field reconstruction is then obtained by sampling
$\mathbfit{v}$ from this probability distribution. Before explaining how
the sampling is done, we turn now our attention to the definitions of priors.

\subsubsection{Priors}\label{sec:priors}
We assume uniform priors on $h_{\text{eff}}$ and
$\sigma_{\text{NL}}$ (see Tab.~\ref{tab:summary}). Following the model described in
Section~\ref{sec:def-not}, we take into account the correlations between the
peculiar velocities by
adopting the following prior for $\delta$:
\begin{equation}
  \label{eq:20}
  \mathcal{P}(\hat{\delta}) = \prod_{\mathbfit{k}} \frac{1}{\sqrt{2\pi P(k)}}
  \exp \left(-\frac{|\hat{\delta}(\mathbfit{k})|^{2}}{2 P(k)}\right)
\end{equation}
where $P(k)$ is the power spectrum and has been defined in 
Eq.~\ref{eq:deltacorr}.
The corresponding prior on the linear velocity field is:
\begin{equation}
  \label{eq:priorv}
  \mathcal{P}(\mathbfit{v}) =
  |2\pi \Psi_{\alpha,\beta} |^{-1/2} \exp \left(-v_{\alpha} \Psi^{-1}_{\alpha,\beta} v_{\beta}\right)
\end{equation}
where $(\alpha,\beta)$ denoted cartesian components and
  $\Psi_{\alpha,\beta}$
  is the velocity-velocity correlation tensor and is defined in the Appendix~\ref{sec:line-pecul-veloc}.\\

Because of the aforementioned Malmquist biases, the prior on the distances can play a significant role
in the analysis. We use two types of empirical priors. 
The first one is a piecewise normal distribution defined by:
\begin{equation}
  \label{eq:prior_norm}
      \mathcal{P}^{(1)}\left(d_{i} | a,b,c\right) =\frac{1}{\sqrt{2
        \pi}(b+c)} \left\lbrace
\begin{array}{ll}
\exp \left( -\frac12 \frac{(d_{i}-a)^{2}}{b^{2}}\right)  &  \mbox{if
                                                               $d_{i}
                                                               \leq a$ }\\
\exp \left( -\frac12 \frac{(d_{i}-a)^{2}}{c^{2}}\right) &  \mbox{otherwise}
\end{array}
\right.
\end{equation}
where $(a,b,c)$ are the shape parameters of the
  function.

The
second one is a power-law with an exponential cutoff, as proposed by \citet{2016MNRAS.457..172L} :
\begin{equation}
  \label{eq:emp}
      \mathcal{P}^{(2)}\left(d_{i} | a,b,c\right) = 
      \frac{1}{N(a,b,c)}\left(d_{i}\right)^{a} \exp \left(-\left(d_{i}/b \right)^{c} \right).
\end{equation}
 The normalization factor $N(a,b,c)$ is non-analytical and is
computed numerically.

These two priors have the properties that we expect for distance priors:
they are bell-shaped curves allowing an asymmetry and with an exponential
cutoff at large distance. The shape parameters $(a,b,c)$
allow for some flexibility of these priors and determine the mean
value, standard deviation and skewness of the distributions. Since we
are not able to establish the selection function(s) of CF3, we use an
empirical approach and choose to fit the shape parameters together with
the other parameters of the current model. Leaving these parameters free allow
to take into account the volume and selection effects while not imposing
strong constraints on derived luminosity distances.
In practice, we attribute the prior $\mathcal{P}^{(2)}$
to distances of subsamples (ii), (iii) and (iv) described in
Section~\ref{sec:selec}, and 
$\mathcal{P}^{(1)}$ to distances of the 6dF subsample. Distances of subsample (v) are
given a uniform prior. We will see in Section~\ref{sec:results} that
they describe correctly the posterior distribution of distances and in
Section~\ref{sec:hubble-constant} the effect of changing the prior functions.
The model parameters are summarized in Table~\ref{tab:summary}.
\begin{table*}
  \centering
  \caption{Observations and parameters used in the present model. For
    the observations, we specify the number and errors. For the
    parameters, we specify the corresponding priors.}
  \begin{tabular}{|c|p{4cm}|c|p{6cm}|}
    \hline
    Fixed parameters & Description & Number & Notes and priors\\
    \hline
    $N$ & Number of galaxy and groups &  1 & $N=11483$\\
    $L$ & Length of the box side &  1 & $L=800$ Mpc/$h_{75}$\\
    $M$ & One dimensional size of the grid &  1 & $M=128$\\
    $\mu$ & Distance moduli & N & Normally distributed with 
                                  standard deviation $\sigma_{\mu}$. \\
    $\sigma_{\mu}$ & Errors on distance moduli & N &  \\
    $z$ & Observed redshift & N & Normally distributed with
                                  standard deviation
                                  $\sigma_{z}$\\
    $\sigma_{z}$ & Error on the observed redshifts & 1
                                            &$\sigma_{z}=\frac{\sigma_{cz}}{c}=50$
    km/s/$c$\\
    \hline
\hline
    Free parameters & &  & \\
\hline
$\mathbfit{v}$ & The linear velocity field sampled on a
                              grid & $M^{3}$ & $\Lambda$CDM prior
    defined by Eq.~\ref{eq:priorv}\\
$d$ & Luminosity distances & $N$ & Empirical priors defined by
                                       Eqs.~\ref{eq:prior_norm}
                                       and~\ref{eq:emp} and/or uniform, depending on the
                                       membership in the subsamples
                                       defined in
                                       Section~\ref{sec:groups}.\\
$h_{\text{eff}}$ & Effective shift of the distance moduli scale&1 &
                                                                    Uniform
                                                                    prior
    within $[0.5,1.5]$\\
$\sigma_{\text{NL}}$ & Gaussian standard deviation modeling the
                       departure from linearity&1 &Uniform prior within $[50,1500]$ km/s\\
$(a,b,c)$ & Hyperparameters defining the distance priors shapes & $3
                                                                  \times
                                            4$&
                                                                      Uniform
                                                                      priors
    depending on the prior function\\
\hline
  \end{tabular}

  \label{tab:summary}
\end{table*}

\subsection{Sampling}
\label{sec:sampling}
After we have constructed the likelihood and the priors of the model, we
need to sample $\mathbfit{v}$ from the posterior distribution given by
Eq.~\eqref{eq:bayes}. In this section, we briefly explain how this
sampling is done by a
Markov Chain Monte Carlo (MCMC) method. For
more technical details, the reader is referred to
\citet{2016MNRAS.457..172L}. The sampling method is the partially
collapsed blocked Gibbs sampling algorithm. Gibbs sampling is a MCMC
method where each parameter is drawn from its conditional probability
given the other parameters. Schematically, if we want to sample $n$
parameters $\{x_{i}\}$, the sampling will be done using the following
scheme down to the Markov step $j$:
\begin{eqnarray*}
  \label{eq:Gibbs}
  x^{1}_{0} &\leftarrow& \mathcal{P}\left(x_{0}| \{x^{0}_{i}\}_{i \in
      [1:n]}\right)\\
&\cdots&\\
  x^{1}_{i} &\leftarrow& \mathcal{P}\left(x_{i}| \{x^{1}_{i}\}_{i \in
      [1:i-1]}, \{x^{0}_{i}\}_{i \in
      [i+1:n]}\right)\\
&\cdots&\\
x^{j}_{i} &\leftarrow& \mathcal{P}\left(x_{i}| \{x^{j}_{i}\}_{i \in
      [1:i-1]}, \{x^{j-1}_{i}\}_{i \in
      [i+1:n]}\right)\\
&\cdots&
\end{eqnarray*}
Our parameter space is:
\begin{equation}
  \label{eq:params}
  \Theta = \lbrace h_{\text{eff}},
  \mathbfit{v},\sigma_{\text{NL}},\lbrace d_{i} \rbrace,\lbrace a,b,c\rbrace \rbrace
\end{equation}
and we need to compute each conditional probability from
the likelihood. However,
we note that $h_{\text{eff}}$ is strongly correlated with the velocity
field, and we draw this parameter from its conditional probability marginalized over
the velocity field to make the sampling more efficient. This is called collapsed Gibbs sampling. At the end,
our sampling is the following procedure:
\begin{enumerate}
\item We first sample $h_{\text{eff}}$ from the probability distribution
  marginalized over the velocity field:
  \begin{equation}
    \label{eq:heff}
    \mathcal{P}\left(h_{\text{eff}} | d,\sigma_{\text{NL}}\right) =
    \mathcal{N}\left(\mathbfit{v}^{r}(h_{\text{eff}}) ; \mathbfit{0},\mathbfss{C}(h_{\text{eff}}) \right)
  \end{equation}
where $\mathbfit{v}^{r}$ is the vector of the galaxies radial peculiar
velocities and $\mathbfss{C}$ is the velocity autocorrelation matrix
defined in the Appendix~\ref{sec:line-pecul-veloc}. Those two quantities
implicitly depend on the parameters $\sigma_{\text{NL}}$
and $d$.
\item We draw $\sigma_{\text{NL}}$ from the conditional probability:
  \begin{equation}
    \label{eq:psnl}
    \mathcal{P}\left(\sigma_{\text{NL}} | h_{\text{eff}},
      d,\mathbfit{v}\right) = \prod^{N}_{i}
      \mathcal{N}\left( v^{r}_{i}
        ;\mathbfit{v}(\mathbfit{r}_{i})\cdot \hat{r}_{i},\sigma_{z,i}^{2}(1+\bar{z}_{i})^{-2}+ \sigma^{2}_{NL}  \right)
  \end{equation}
\item We draw a constrained realization of
  $\mathbfit{v}$. This is done by using the Hoffman-Ribak
  algorithm~\citep{1991ApJ...380L...5H}. Appendix~\ref{sec:hoff} reminds
  the reader about the algorithm.
\item From the sampled constrained realization, we generate a new set
  of luminosity distances $d$ with the following probability:
  \begin{equation}
    \label{eq:pd}
    \mathcal{P}\left(d| h_{\text{eff}},
      \sigma_{\text{NL}},\mathbfit{v}\right) \propto \mathcal{L}
    \times \mathcal{P}\left(d\right)
  \end{equation}
\item Eventually we fit the hyperparameters $(a,b,c)$ of the distance
  prior functions over the generated distances.
\end{enumerate}
This procedure is carried until convergence. At the end we have a number
of realizations of all parameters following the posterior probability
defined by Eq.~\eqref{eq:bayes}. The reconstructed velocity field is
assumed to be the mean of the linear velocity field samples
$\langle\mathbfit{v}\rangle_{MCMC}$ and the error is the standard
deviation. The method has been
tested on different mocks by \citet{2016MNRAS.457..172L} and we test our
specific implementation as described in
Appendix~\ref{sec:mock}. 

\section{The \emph{Cosmicflows}-3 peculiar velocity field}\label{sec:results}
In this section we present the results of the method applied on
CF3 catalog. We chose to assume
$(\Omega_{m},\Omega_{\Lambda},H_{0}) = (0.3,0.7,75.0)$. We fit the distance
priors by the empirical function defined by Eq.~\ref{eq:emp}
for the subsamples (ii,iii,iv) defined in Section~\ref{sec:selec}. The distances from the subsample
(i) corresponding to 6dF data are
fitted assuming the function described by Eq.~\ref{eq:prior_norm}. We do so
to ensure the possibility for the prior function to model a sharp cut-off
in distances. 
\begin{figure}
  \centering
   \includegraphics[width=1\linewidth]{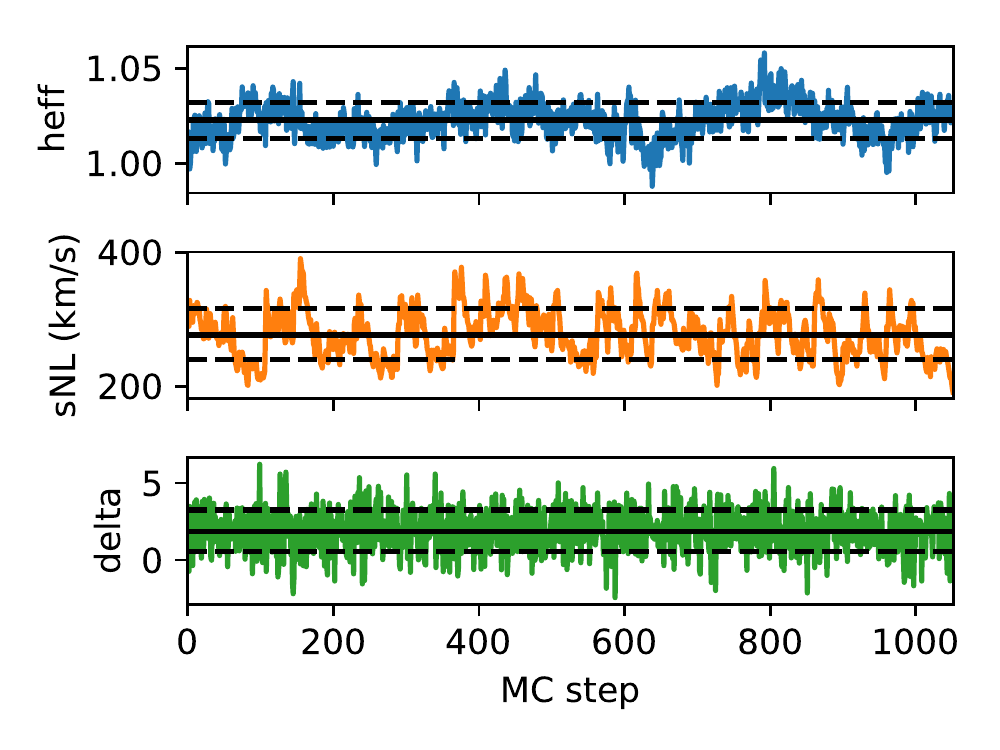}
  \caption{The Monte Carlo Markov Chain for three parameters: (top) 
    the effective reduced Hubble constant $h_{\text{eff}}$; (mid) the non linear
    dispersion $\sigma_{\text{NL}}$; (bottom) the reconstructed overdensity field
    near Virgo, at coordinates $(SGX,SGY,SGZ) =  (-3.6, 15.6, -0.7)$ Mpc/$h_{75}$. The three black lines represent the
    15.9$^\text{th}$ (dashed), 50$^\text{th}$ (plain) and
    84.1$^\text{th}$ (dashed) percentiles.}
  \label{fig:mcmc_1H}
\end{figure}

\begin{figure*}
  \centering
   \includegraphics[width=1\linewidth]{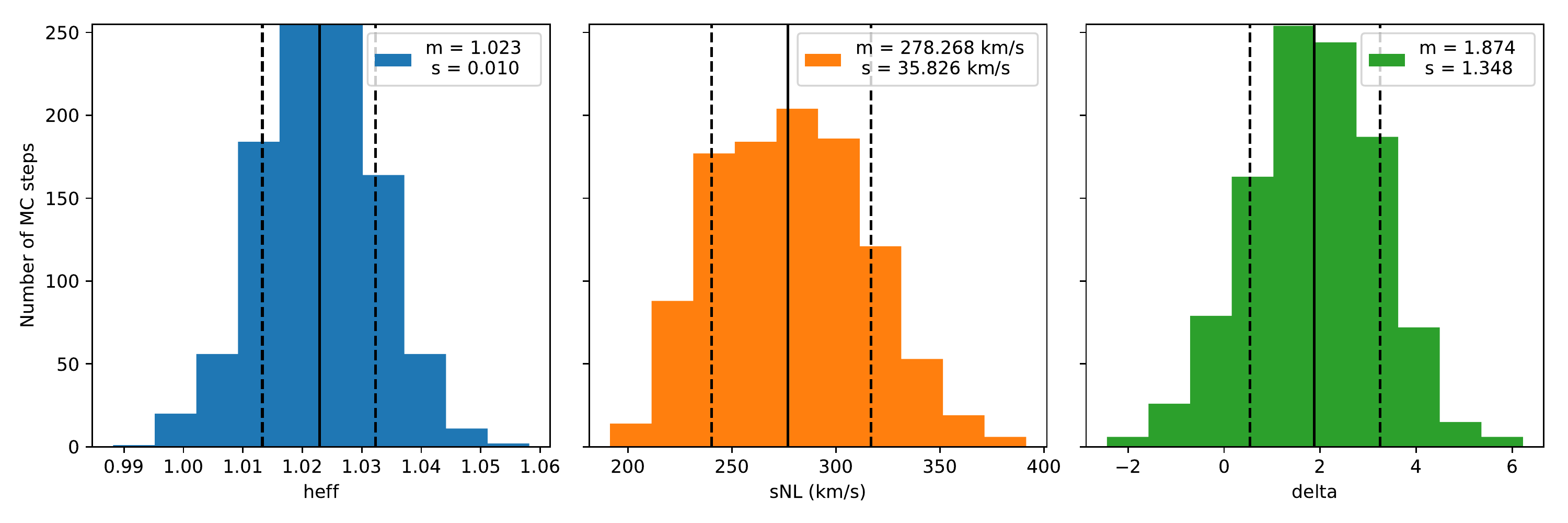}
  \caption{The posterior distribution for three parameters:  (left) 
    the effective reduced Hubble constant $h_{\text{eff}}$; (mid) the non linear
    dispersion $\sigma_{\text{NL}}$; (right) the reconstructed overdensity
    at Virgo, of coordinates $(SGX,SGY,SGZ) =  (-3.6, 15.6, -0.7)$
    Mpc. The value of $m$ given in the legend corresponds to the mean
    value of the posterior, while $s$ corresponds to its standard
    deviation. The three black lines represent the
    15.9$^\text{th}$ (dashed), 50$^\text{th}$ (plain) and
    84.1$^\text{th}$ (dashed) percentiles.}
  \label{fig:hist_1H}
\end{figure*}

\begin{figure}
  \centering
   \includegraphics[width=1\linewidth]{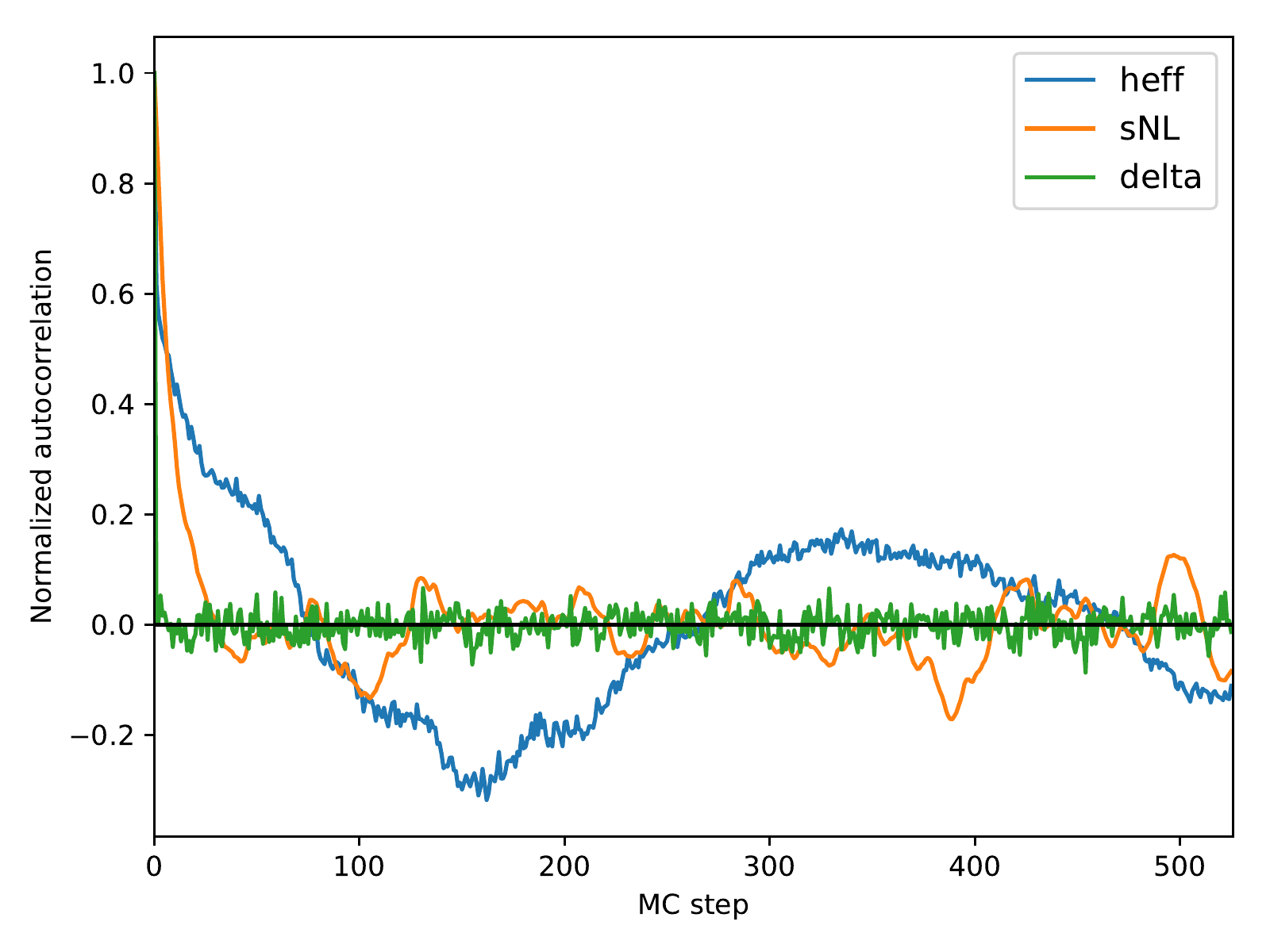}
  \caption{The normalized auto-correlation of the MCMC for three parameters:
    the effective reduced Hubble constant $h_{\text{eff}}$; the non linear
    dispersion $\sigma_{\text{NL}}$ and the reconstructed overdensity
    at Virgo, of coordinates $(SGX,SGY,SGZ) =  (-3.6, 15.6, -0.7)$
    Mpc. The black solid line corresponds to a null auto-correlation.}
  \label{fig:autocorr_1H}
\end{figure}
\begin{figure}
  \centering
   \includegraphics[width=1\linewidth]{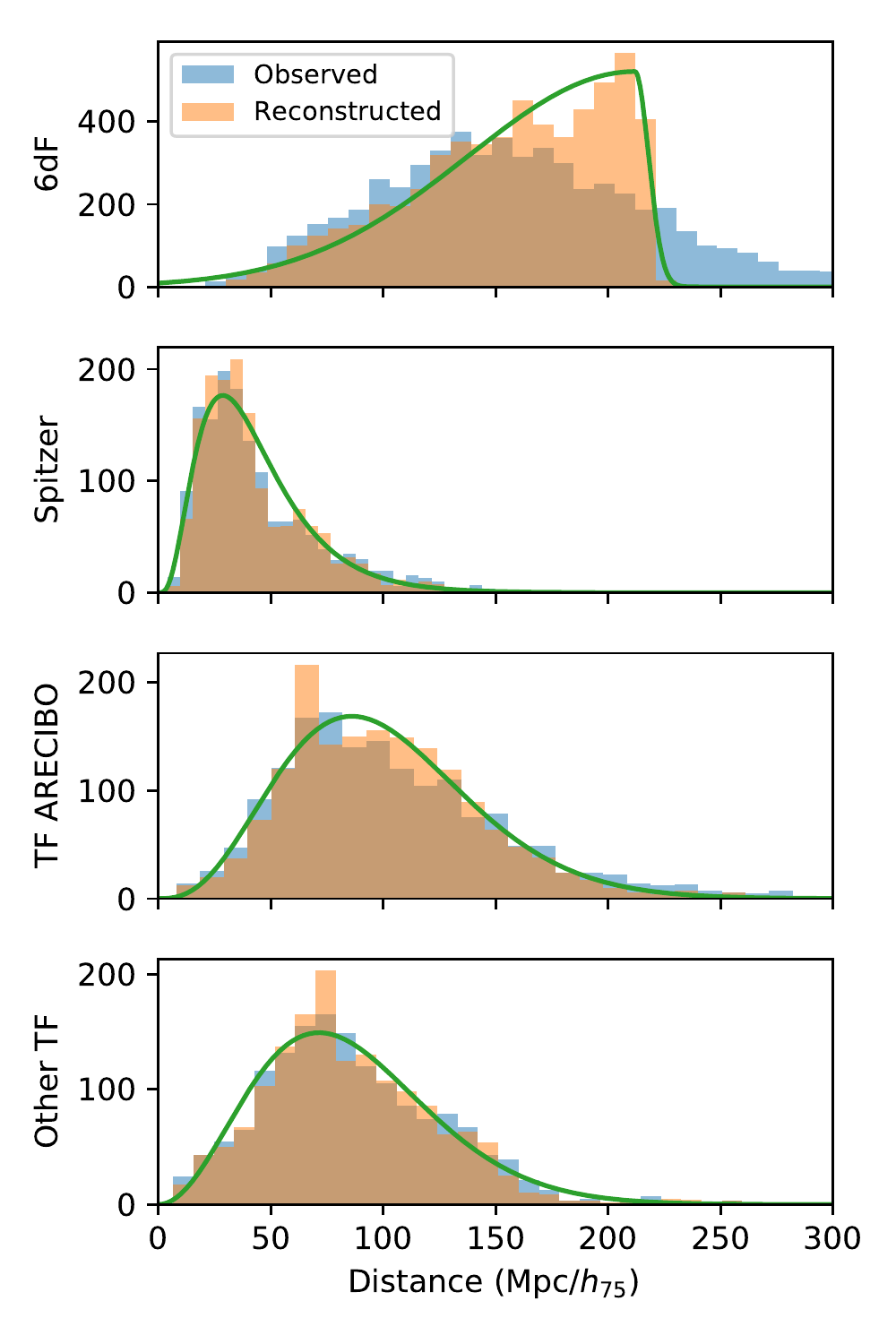}

  \caption{Histogram of measured (blue) and fitted (orange) distances for
  the five subsamples described in Section~\ref{sec:data}. The green lines
are the fitted priors. From top to bottom: (1) 6dF sample,  mixed
Gaussian prior
distribution; (2) Spitzer galaxies,
empirical prior distribution; (3) TF by Arecibo
Telescope, empirical  prior distribution; (4) TF not covered by Arecibo
Telescope, empirical prior distribution; The empirical prior distribution is
described in Eq~\ref{eq:emp}.}
  \label{fig:priors_1H}
\end{figure}

\begin{figure*}
  \centering
   \includegraphics[width=0.8\linewidth]{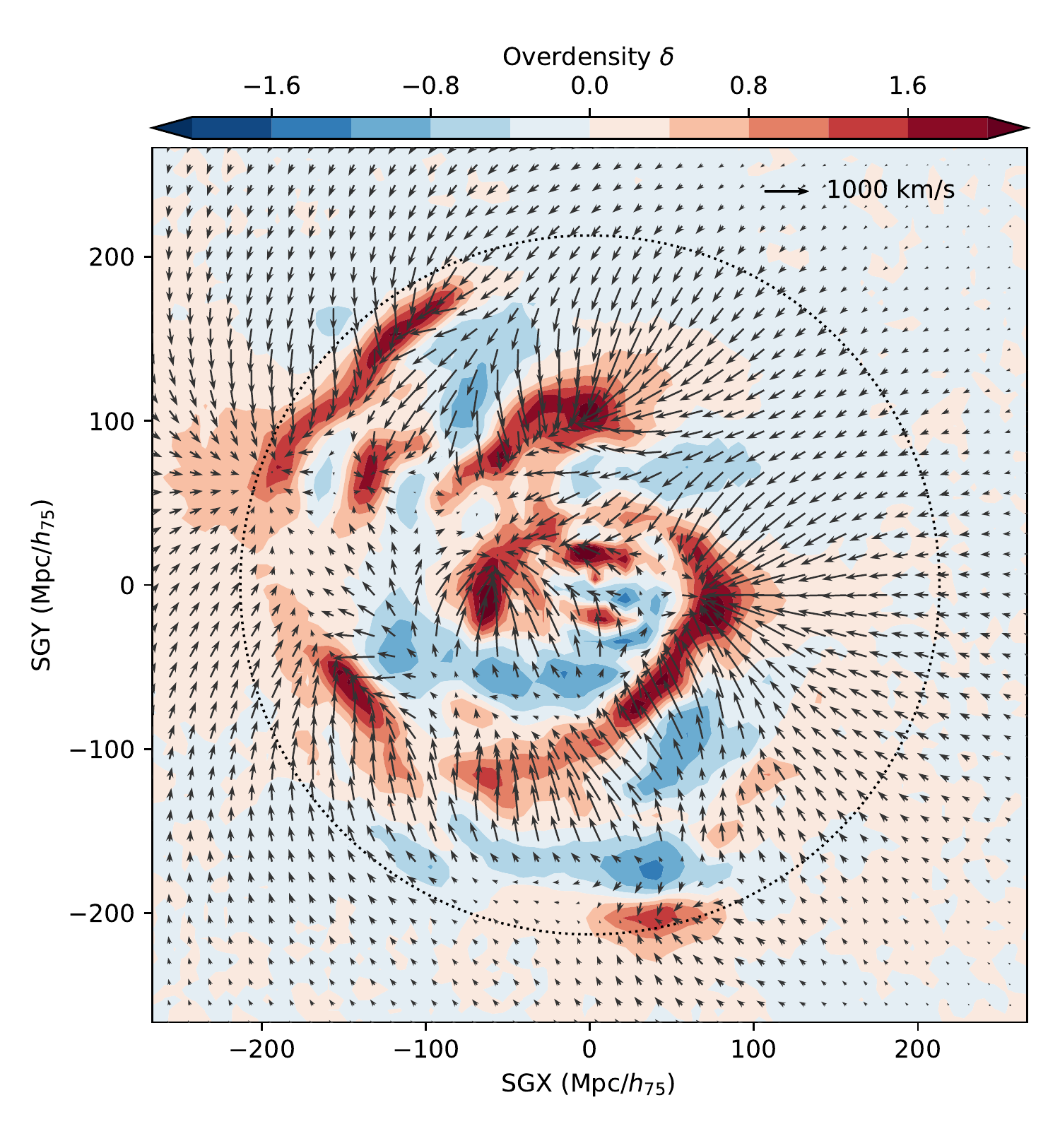}

  \caption{Central part of the CF3 velocity field
    reconstruction in the $SGZ=0$ Mpc/$h_{75}$ slice. The color corresponds to the value of the
    overdensity field while the black arrows represent the
    tridimensional reconstructed linear peculiar velocity field. The
    dotted black
    circle illustrates the edge of the data at $z = 0.054$.}
  \label{fig:delta_1H}
\end{figure*}

\begin{figure*}
  \centering
   \includegraphics[width=1\linewidth]{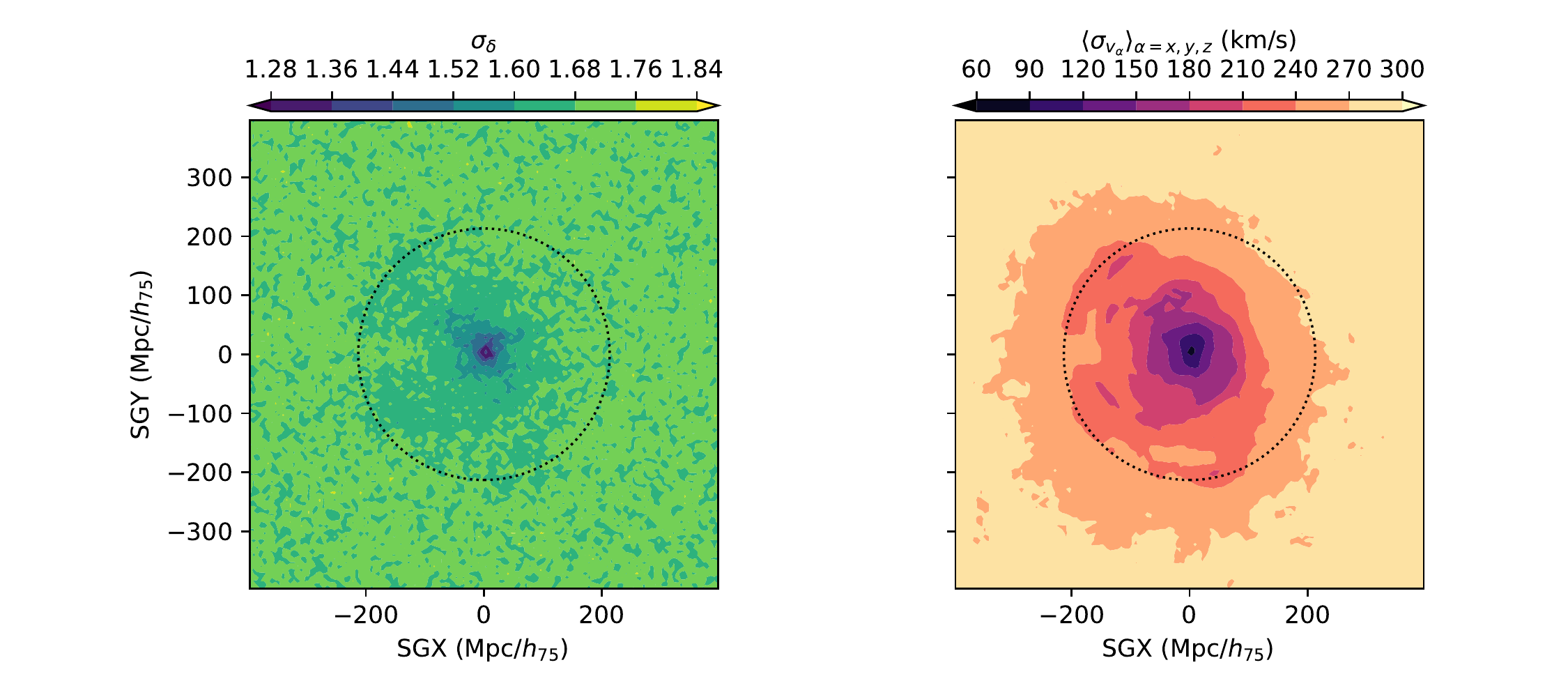}

  \caption{Statistical standard deviation for the sampled overdensity (left) and
    velocity (right) fields in the $SGZ=0$ Mpc/$h_{75}$ slice over the
    full reconstructed box. The standard deviation of the velocity field
    takes into account all three cartesian components. The black
    circle represents the edge of the data at $z = 0.054$.}
  \label{fig:SNR_1H}
\end{figure*}
We fit for a global effective zero-point shift
$h_{\text{eff}}$ and assume the inter-calibration of distances determined
by~\cite{2016AJ....152...50T} and the CMB frame transformations given
by~\cite{2008ApJ...676..184T}. Figure~\ref{fig:mcmc_1H} shows the
resulting MCMC chain for the two parameters
$\sigma_{\text{NL}}$, $h_{\text{eff}}$ and for the overdensity field
reconstructed near the Virgo Cluster at location
$(SGX,SGY,SGZ)=(-3.6,15.6,-0.7)$ Mpc/$h_{75}$ and the corresponding
histograms are given in Fig.~\ref{fig:hist_1H}. We see from these two
figures that the chain has globally converges and result in
approximately Gaussian posterior distributions. 
To further evaluate the convergence of our Markov
Chain, we plot in Fig.~\ref{fig:autocorr_1H} the normalized autocorrelation of
the chains for the two parameters $h_{\text{eff}}$ and
$\sigma_{\text{NL}}$. The autocorrelation of a parameter $f$ for a
correlation length $\tau$ is defined by:
\begin{equation}
  \label{eq:autocorr}
  c_{j}(\tau) = \frac{1}{N-\tau}\sum_{i=1}^{N-\tau}(f_{i}-\bar{f})(f_{i+\tau}-\bar{f})
\end{equation}
where $\bar{f}$ is the mean value of $f$ computed on the $N$ samples. The first decorrelation for a chain corresponds to
the intersection with zero. We can see that the $h_{\text{eff}}$
has a decorrelation length of about $\sim 80$ and the non-linear
dispersion around $\sim 20$ steps. This suggests that over our 1400 MC
steps, there is some 20 independent samples. For each parameter, the
error on the mean of the distribution decreases, as
$1/\sqrt{N_{\text{sample}}}$ after convergence and running the chain further will reduce
this statistical error. We estimate here that the sampling error on the mean
overdensity field (typically $\sim 0.1$) is sufficient considering
the computation cost of the sampling.

We find $h_{\text{eff}}
= 1.02 \pm 0.01$, which suggests that the calibration of CF3 data is
compatible with the assumed fiducial Hubble constant $H_{0} =
75$ km/s/Mpc within $2 \%$. The non-linear dispersion parameter
$\sigma_{\text{NL}}$, which models our lack of knowledge about the non-linear
part of the velocity field, is found at $\sigma_{\text{NL}}=280 \pm 35$ km/s. The fitted value of around 300 km/s appears
to be high compared to the typical value of $100-200$ km/s. This can be due to
underestimation of distance modulus errors, or redshifts errors which
were set here at $\sigma_{cz}=50$ km/s for every galaxy. Also, this
value depends on the efficiency of the grouping described
Section~\ref{sec:groups} at removing entirely the non-linearities in
groups of galaxies. 
Another possibility is that the trilinear interpolation
used to compute the reconstructed radial velocity at the position of
each galaxy artificially increases the departure from linearity modeled
by $\sigma_{\text{NL}}$. Overall this parameter absorbs uncertainties of our
model.

Figure~\ref{fig:priors_1H} shows the fitted priors on the
distances. Overall the shape of the priors is in agreement with the
underlying distance distribution. The case of 6dF data might look
surprising because there is a discrepancy between the fitted distances and
the original ones. However, by considering the distribution in redshifts shown
in Figure~\ref{fig:z}, we can see that the tail of the measured distance
distribution is only due to observational errors and results from the
convolution of the real distance distribution with the Gaussian of
errors. This particular case illustrates the importance of imposing
priors to model selection effects. 

Eventually, the $SGZ=0$ Mpc/$h_{75}$ slice
of the reconstruction is
shown Figure~\ref{fig:delta_1H}. In this figure, the colormap corresponds
to the reconstructed overdensity field and the black arrows to the
tridimensional linear velocity field. We stress that the reconstructed overdensity
and velocity fields are not computed one from another using Eq.~\ref{eq:vfourier}, but rather
sampled simultaneously\footnote{For each MC step, we draw both a constrained
realization of the velocity field and overdensity field from a common
random realization with Eqs.~\ref{eq:9} and~\ref{eq:10}}. The original
field was computed in a box of width 800 Mpc/$h_{75}$, and we show here the
central 500 Mpc/$h_{75}$ where the signal over noise is non-zero. We
can see that the overdensity goes toward zero near the edges, because
there are no observed galaxies at large distances. By eye, we can
identify several cosmic structures that we will shortly list in Section~\ref{sec:cosmography}. We also qualitatively see the match between the overdense
structures and the velocity infalls. These fields
come with statistical errors that can be computed from the resulting
Markov Chain. The standard deviation of the reconstructed fields
$\delta$ and $v^{r}$
are plotted in
Fig.~\ref{fig:SNR_1H}. At large distance, we recover the $\Lambda$CDM
standard deviation of the overdensity and velocity field. In particular,
we recover the value of 300 km/s deviation for the velocity field. This is
due to the absence of data points beyond $z=0.054$, a limit which is
illustrated by the black circle. We can see that the radial
peculiar velocity field seems less noisy than the overdensity field. 
The reason is that the velocity field is more correlated
  than the overdensity as can be seen with
Eq.~\ref{eq:6}. Consequently, the
  root mean square value of the posterior distribution does not capture entirely the
  correlated errors of the velocity field, and these errors correlate on
larger scales.
In the next section, we study the reconstructed overdensity field shown in the
left panel of Fig.~\ref{fig:delta_1H} and compare it with the
actual distribution of galaxies to check the consistency with other observables.

\section{Cosmography overview}
\label{sec:cosmography}
\begin{figure*}
  \centering
   \includegraphics[width=1\linewidth]{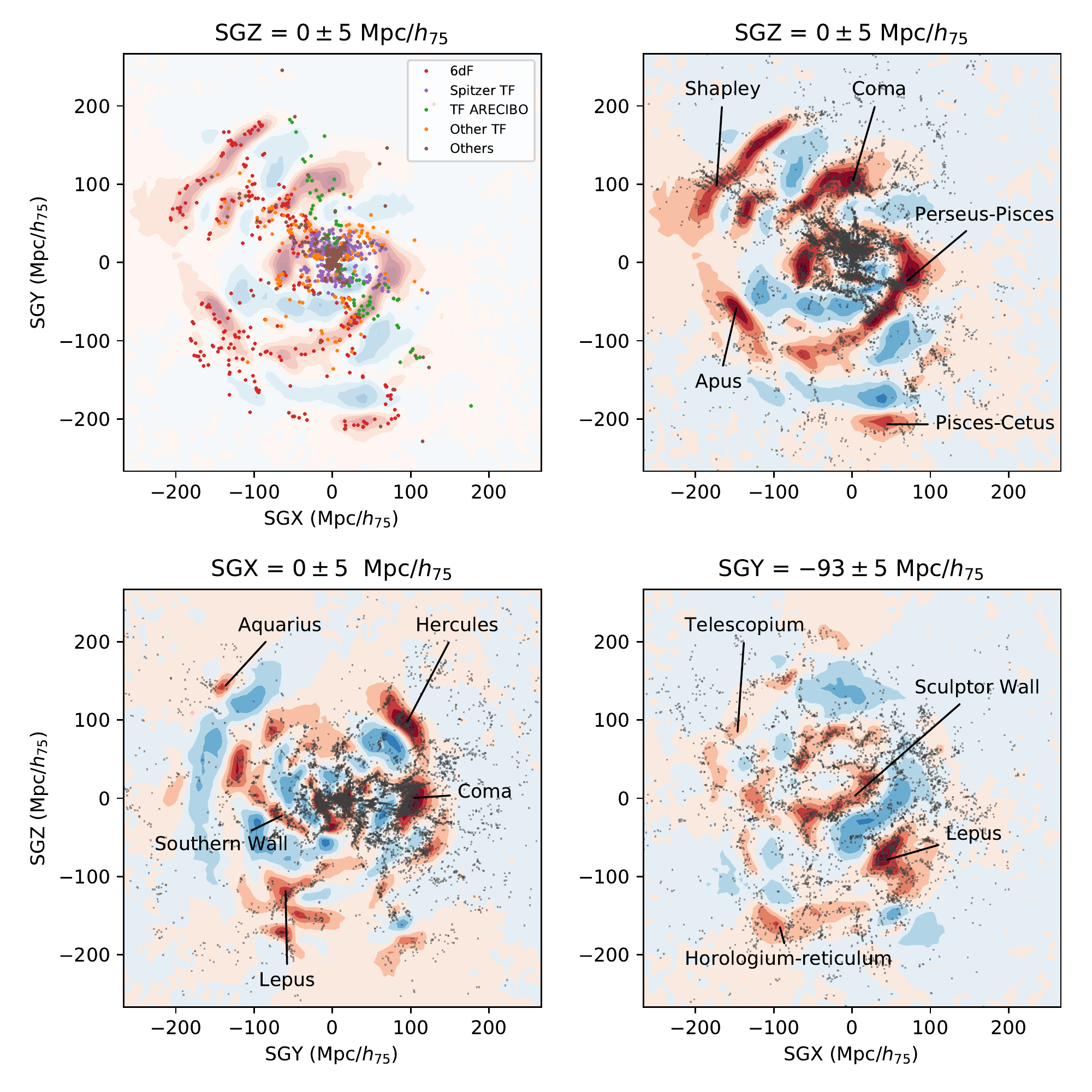}
  \caption{Four slices of the reconstructed overdensity field of
    CF3. The colored dots represent galaxies from CF3 catalog. The color
    depends on the appartenance of galaxies to subsamples defined in
    Section~\ref{sec:selec} and are coded the same way as in Fig.~\ref{fig:z}: (red) 6dF galaxies, subsample (i); (purple) Spitzer TF
    data, subsample (ii); (green) Arecibo TF
    data, subsample (iii); (orange) Other TF
    data, subsample (iv); (brown) Remaining individual and groups
    data, subsample (v).  On other slices, the dark dot represent galaxies from LEDA located at their
    redshift position.  Coordinates of slices: top-left and
    top-right $SGZ = 0$ Mpc/$h_{75}$; Bottom-left $SGX=0$ Mpc/$h_{75}$;
    Bottom-right: $SGY=-93$ Mpc/$h_{75}$. We denominate major structures
  on the different shown slices.}
  \label{fig:slices}
\end{figure*}

In this section we show partial results on the reconstructed overdensity
field. A full description will be made in an upcoming cosmography
paper. Main results are presented in
Figure~\ref{fig:slices} where we plot the overdensity field in
three different slices corresponding to coordinates $SGZ=0$ Mpc/$h_{75}$ (twice), $SGX=0$ Mpc/$h_{75}$ and $SGY=-93$ Mpc/$h_{75}$ going from top
left to bottom right. We first look at the two top panels. On the left,
the colored dots represent CF3 data within a slab of 10 Mpc/$h_{75}$ thickness, while on
the right the dark dots are all available galaxies of the LEDA
database located at their redshift positions within the same depth. The
colored galaxies show the anisotropy of CF3 catalog. The southern hemisphere
is mainly covered by 6dF galaxies (in red), while there are fewer
data points in the northern  hemisphere. Looking at the right panel, we
can observe a good agreement between the reconstructed overdensity
field and the location of galaxies. At large distances, we notice small
discrepancies (for example around $SGY = -210$ Mpc/$h_{75}$ (top right panel)). The
reconstructed structures seem to be shifted compared to the
redshift positions. This issue is tightly linked to the recovered value of
$h_{\text{eff}}$, which is prone to systematics. This might be a hint of a shift between zero-point
calibrations depending on the region covered. We discuss this issue in
Section~\ref{sec:hubble-constant}. We add to the plots the name of some
known structures, such as Coma, Shapley, Perseus-Pisces, Apus and
Pisces-Cetus. The two bottom panels correspond to other slices of the
Local Universe reconstruction. Again, there is agreement between the reconstructed
overdense regions and the galaxy distribution. The bottom right panel
exhibits some distant structures  which are less known, such as the
Southern Wall, Telescopium or Lepus. 

\section{Discussion}
\label{sec:discussion}
In this paper, we described our linear peculiar
velocity field reconstruction method. We then applied it to the CF3 catalog, considering generic modeling of the
prior distance distributions and potential shift on the zero-point
calibration of all distances.  We saw in Section~\ref{sec:cosmography}
how this reconstruction can be used to study our local environment and
the unbiased distribution of dark matter in the $\Lambda$CDM framework. We now turn to the discussion
of the limits and
possible improvements of the presented method.

\subsection{The Hubble Constant}
\label{sec:hubble-constant}

We introduced an effective reduced Hubble constant
$h_{\text{eff}}$ to absorb uncertainty on the calibration of the
distance indicators and assumed a Hubble constant of $H_{0} = 75$
km/s/Mpc. It is worth noticing that the value of the
effective Hubble constant is correlated with our choice of distance priors. This
systematic is studied in Appendix~\ref{sec:chang-dist-priors} on a
subsample of the CF3 catalog. We find that imposing uniform priors on the
distances lowers the value of $h_{\text{eff}}$ by around $5 \%$
(equivalent of around 3 km/s/Mpc). This suggests that the parameter
 $h_{\text{eff}}$ not only absorb uncertainties on the assumed Hubble
 Constant, but also compensate for residual homogeneous Malmquist bias
 that is not modeled by the simple shape of our distance priors. 

The CF3 catalog contains distances from six distinct methodologies.
With each of these methodologies there are contributions from multiple sources.
It is fundamentally important that the distances from the diverse inputs be consistent 
in zero point and free of systematics with redshift.  An important feature in the construction of CF3
was large overlaps between sources.  The complex interlacing is discussed by \citet{2016AJ....152...50T}.
The grounding assumption was all measurements of the distance to a galaxy or a cluster of galaxies 
should give the same value on average.  The scale is bootstrapped from fundamental Cepheid
and RR Lyrae calibrators in our galaxy.  Consequently the CF3 product is distances in Mpc derived 
independently from knowledge of velocities.

There is one relatively weak link.  The 6dF sample is a major component of CF3 that explores 
a domain that is poorly covered by other samples. The scale linkage is established through 
84 individual galaxies with alternate measures and 381 groups with distances to members
from alternate methods.  Given the importance of the 6dF contribution, in Appendix~\ref{sec:2H}
we consider the possibility of 
a scale mismatch by introducing an independent $h_{\text{eff}}$ parameter associated with the 6dF contribution.
A small difference is found between the optimal values for the 6dF component and the rest of 
the CF3 contributions, namely the effective Hubble constant is reduced
by 3\% for the 6DF sample. We find that the reconstructed
radial velocity field is affected by the introduction of this additional parameter, and that
the non-linear dispersion parameter $\sigma_{\text{NL}}$ is
reduced by 20\%.  It is not clear, though, that the addition of a parameter has improved the model.
The additional $h_{\text{eff}}$ parameter for the 6dF component could mask velocity streaming specific to 
the sector uniquely sampled by 6dF.  Our primary interest is to map the velocity field
so for that purpose the conservative assumption is to not add an additional parameter.
Rather, we assume that the 6dF distances were properly linked to the 
other CF3 components, and model departures from the fiducial $H_0$ are described by a single value of $h_{\text{eff}}$.

\subsection{Non-linearities}
\label{sec:non-linearities}
The modeling of non-linearities can also be a source of systematics of our
reconstruction. Our treatment involves the grouping described in
Section~\ref{sec:groups}. The merging of data within groups has clear advantages.  While what we directly observe are individual galaxies,
the test particles we want to follow in a linear reconstruction are collapsed haloes.  If a halo contains multiple galaxies 
then we do best to average over the constituent properties.  The groups are formulated from a large redshift catalog \citep{2015AJ....149..171T} and we average over the redshifts of all associated members.  Then we find weighted averages of the distances from 
those group members in CF3.  Distances of groups with many contributions can have low formal errors.

 The remaining non
  linearities (outside of the groups) at redshift zero are modeled by
  a unique Gaussian dispersion to the observed velocities. The width
  of this distribution, namely $\sigma_{\text{NL}}$, is fitted with the other
  parameters. This is an approximation: non-linear motions are
  correlated down to $k \gtrsim 0.1$ Mpc$^{-1}$ scale and so are not
  randomly independent. However, including non-linear correlations within the actual method is a
  technical challenge since the correlations in the non-linear regime
  are not isotropic anymore. An interesting alternative is proposed by~\citet{2018NatAs.tmp...91H} who try to use
  constrained simulations in order to reconstruct the non-linear part of
  the peculiar velocity field. 
Other options could also be tried like fixing the $\sigma_{\text{NL}}$ parameter
  at a value constrained by $\Lambda$CDM non-linear simulation, or
  adding a discrete parameter for each galaxies to model their association
  to non-linear regions, such as
  in~\citet{2016MNRAS.457..172L}. The advantage of the latter method is
  to model both non-linearities and possible outliers.

\subsection{Selection effects}
\label{sec:select-effect}
Selection effects on distances play a considerable role in peculiar velocity
analyses even though less critical than in galaxy number counts analysis. We described in Section~\ref{sec:malmquist-bias-1} the
Malmquist biases appearing when the selection
effects are not properly modeled. In our Bayesian approach, the
selection effects are modeled by constructing the probability $\mathcal{P}(d)$ of a
galaxy having a luminosity distance $d$ knowing that we observed it. Our
model suppose that this probability only depends on the actual distance
$d$ and fitted hyperparameters ($a$, $b$, $c$). In particular,
individual CF3 distances are unbiased~\citep{2016AJ....152...50T}, and
the probability $\mathcal{P}(d)$ does not depend on the observed distance modulus
$\mu$. 

We note however that selection effects would need further investigation
in future works on peculiar velocities. As suggested
by~\cite{2017arXiv170603856H}, a proper model needs the introduction of the generalized likelihood
$\mathcal{L}^{\prime}$:
\begin{equation}
  \label{eq:Pselec}
  \mathcal{L}^{\prime} = \mathcal{L} \times 
  \frac{\mathcal{P}(\text{selection}|\text{data},\text{parameters})}{\int
    \mathcal{P}(\text{selection}|D,\text{parameters})\mathcal{P}(D|\text{parameters})\mathrm{d}
  D}
\end{equation}
where the integration is over all possible observational data $D$. Since the selection
function of CF3 is not analytical it is challenging to write the term
$\mathcal{P}(\text{selection}|D,\text{parameters})$ representing the
probability of selecting a galaxy measurement given its overall
properties and the model's free parameters. Sampling from the extended
likelihood $\mathcal{L}^{\prime}$ is consequently hard and unpractical
for the current analysis.

Our approach, while it is approximate, is robust and secure since we fit the priors
directly on the reconstructed distances. Doing so, we value CF3 distances
over eventual prior information. Also, it is worth noticing that the CF3 catalog benefits
from the multiplicity of methods included in it. The selection effects being
different from one methodology to another, the overall reconstructed
velocity field should not be subject to individual method specificity.
As mentionned in Section~\ref{sec:hubble-constant}, we tested the effect
of changing the priors on the distances and the results are presented in Appendix~\ref{sec:chang-dist-priors}.

\section{Conclusions}
\label{sec:conclusion}
This article presents an algorithm to reconstruct the
linear peculiar velocity field up to $z \sim 0.054$ from the \emph{Cosmicflows-3}
catalog. We have been able to reconstruct for the first time the Local
Universe velocity field from \emph{Cosmicflows-3} data, and showed some results in
terms of cosmography. We also associated a corresponding map
of statistical errors on both the overdensity and velocity field. The reconstructed field will
be used for both cosmological analysis and
cosmography. We have stressed the limits of the current method, especially
about the model of non-linearities and selection effects. We showed that
these effects were not completely negligible and should be more precisely modeled
in future analyses.

Above all, this article highlights the ability of peculiar velocities
 to probe the matter distribution. With only about 11,000 tracers, we were able to map and
identify overdense and underdense regions in the Local Universe, and
showed the good agreement with redshift surveys containing more than 150,000
galaxies. As suggested by \citet{2016MNRAS.457..172L}, the model could be extended to estimate
cosmological parameters such as the Hubble constant $H_{0}$ or the
growth rate $f\sigma_{8}$. With upcoming large distance datasets, coming
from TAIPAN (\citet{2017PASA...34...47D}, $\sim 50,000$ FP distances), WALLABY (\citet{2012MNRAS.426.3385D}, $\sim 60,000$
TF distances) and LSST (\citet{2018arXiv181200515L}, $\sim 100,000$ SNIa
distances), peculiar velocity analyses will
need an accurate model to avoid systematics in the
determination of cosmological parameters. The method presented here is
to be considered as a baseline of such model. 

\section*{Acknowledgements}

R.G. thanks Mickael Rigault for fruitful discussions. We are grateful to the colleagues
maintaining LEDA Lyon extragalactic database Dmitry Makarov and Philippe
Prugniel. This research was supported by Institut Universitaire de
France and CNES. This project has received funding from the European Research Council (ERC) under the European Union's Horizon 2020 research and innovation programme (grant agreement No 759194 - USNAC).



\bibliographystyle{mnras}
\bibliography{bibli} 



\appendix

\section{Linear peculiar velocity distribution in the $\Lambda$CDM
  model}
\label{sec:line-pecul-veloc}
Our framework is the linear theory of structure formation within a flat
$\Lambda$CDM model. The initial perturbations of such Universe are
characterized by their Gaussian statistics as observed from the Cosmic Microwave Background
\citep{2015arXiv150201589P}. In the linear theory, the Fourier modes of
these perturbations has grown independently and the
overdensity field of our Local Universe can still be described by its power
spectrum $P(k)$:
\begin{equation}
  \label{eq:4}
\langle \delta(\mathbfit{k}) \delta^{*}(\mathbfit{k}')\rangle = (2 \pi)^{3}
   \delta_{D}(\mathbfit{k} - \mathbfit{k}') P(k)
\end{equation}
where $\delta(\mathbfit{k})$ is the Fourier transform of
$\delta(\mathbfit{r})$. Such a linear approximation is only valid at
large scales, \emph{i.e.} $k\lessapprox 0.2~h.\text{Mpc}^{-1}$, and consequently the present
analysis aims at recovering only the very large scale structures, down
to scales of few tens of Mpc. 
Given the cosmological growth rate $f$ ($f$ depends on the adopted
cosmology), one can compute the
velocity field $\mathbfit{v}(\mathbfit{r})$ by:
\begin{equation}
  \label{eq:5}
  \nabla \cdot \mathbfit{v} = - H_0 f \delta
\end{equation}
which can be written in Fourier space as:
\begin{equation}
  \label{eq:6}
  \mathbfit{v}(\mathbfit{k}) = i H_0 f\frac{\mathbfit{k}}{k^{2}} \delta(\mathbfit{k})
\end{equation}
Consequently, the velocity-velocity two point correlation tensor is, in
configuration space:
\begin{equation}
  \label{eq:vvcorr}
  \Psi_{\alpha,\beta}(\mathbfit{r}) \triangleq \langle  \mathbfit{v}(\mathbfit{r'}) \mathbfit{v}(\mathbfit{r'}+
  \mathbfit{r})\rangle_{\alpha \beta} = \frac{(H_0 f)^{2}}{(2 \pi)^{3}}
  \int_{0}^{\infty} \frac{k_{\alpha}k_{\beta}}{k^{4}} P(k) e^{-i
    \mathbfit{k}\cdot\mathbfit{r}} \dd \mathbfit{k}
\end{equation}
In practice, $\Psi_{\alpha,\beta}(\mathbfit{r}) $ can be expressed using the radial
and transverse correlations functions $\psi_{R}$ and $\psi_{T}$~\citep{1989ApJ...344....1G}:
\begin{equation}
  \label{eq:psi}
  \Psi_{\alpha,\beta}(r) = \psi_{T}(r) \delta^{K}_{\alpha,\beta} +
  \left(\psi_{R}(r) - \psi_{T}(r)\right)\hat{r}_{\alpha}\hat{r}_{\beta}
\end{equation}
\begin{equation}
  \label{eq:psiR}
  \psi_{R}(x) \triangleq \frac{1}{2\pi^{2}}\int_{0}^{\infty} \left(j_{0}(kx) -
    \frac{2 j_{1}(kx)}{kx} \right) P(k) \mathrm{d}k
\end{equation}
\begin{equation}
  \label{eq:psiT}
  \psi_{T}(x) \triangleq \frac{1}{2\pi^{2}}\int_{0}^{\infty} 
    \frac{j_{1}(kx)}{kx} P(k) \mathrm{d}k
\end{equation}
where $j_{0}$ and $j_{1}$ are the zero$^{\text{th}}$ and first order spherical Bessel functions
 and $\delta^{K}$ is the Kroenecker delta.
In practice, we use tabulated $\psi_{R}$ and $\psi_{T}$, and use linear
interpolation between the sampled positions.
We also define the covariance matrix $\mathbfss{C}$ of a set of radial peculiar
velocities by:
\begin{equation}
  \label{eq:dvcorr}
  \left[\mathbfss{C}\right]_{i,j} \triangleq \langle v^{r}_{i} v^{r}_{j}
  \rangle +  \langle \epsilon_{i} \epsilon_{j}
  \rangle
  = \sum_{\alpha,\beta}\hat{r}_{i,\alpha} \hat{r}_{j,\beta}
  \Psi_{\alpha,\beta} + \langle \epsilon_{i} \epsilon_{j}
  \rangle
\end{equation}
Usually, the matrix of errors $\langle \epsilon_{i}\epsilon_{j}\rangle $ is
taken as the sum of the error on the redshift measurement plus a
dispersion due to non-linearities at $z\sim 0$, $\sigma_{\text{NL}}$ :
\begin{equation}
  \label{eq:15}
  \langle \epsilon_{i}\epsilon_{j}\rangle  = \delta^{K}_{ij}(\sigma^{2}_{cz} + \sigma^{2}_{\text{NL}})
\end{equation}

\section{The Hoffman-Ribak algorithm}
\label{sec:hoff}
The marginalized probability density for the overdensity field
$\mathcal{P}(\{\delta(\mathbfit{r}_{j})\} |
\{d_{i}\},\sigma_{\text{NL}},h_{\text{eff}})$ is
\begin{equation}
  \label{eq:23}
  \begin{split}
  &\mathcal{P}(\delta |
\{d_{i}\},\sigma_{\text{NL}},h_{\text{eff}}) \propto\\
&\prod_{i} \gauss{v^{r}(z_{i},\bar{z}_{i}(h_{\text{eff}}d_{i}))}{\mathbfit{v}
  \cdot \hat{\mathbfit{r}}_{i}}{\sigma_{cz}^{2}(1+\bar{z}_{i})^{-2}+
  \sigma^{2}_{NL}} \\
&\times \prod_{j} \frac{1}{\sqrt{2\pi P(k_{j})}}
  \exp \left(-\frac{|\hat{\delta}(\mathbfit{k}_{j})|^{2}}{2 P(k_{j})}\right)
\end{split}
\end{equation}
To sample $\delta$ from this probability density function
\citet{1991ApJ...380L...5H} proposes the following. From the power
spectrum $P(k)$ is generated a random realization $\delta^{RR}$. Then, a
constrained realization is computed with the previous random part and a
correlated one:
\begin{equation}
  \label{eq:9}
  \delta^{CR} = \delta^{RR} + \langle \delta c_{i}\rangle \langle c_{i}c_{j}\rangle ^{-1}
(c_{i}-c^{RR}_{i})
\end{equation}
where $c_{i}$ is a constraint on the sampled
field, here the radial peculiar velocities
$v^{r}$. The matrix $\langle c_{i}c_{j}\rangle =\mathbfss{C}$ is defined in
Appendix~\ref{sec:line-pecul-veloc}, and the correlation between the
overdensity field and the radial peculiar velocity is:
\begin{equation}
  \label{eq:7}
  \langle \delta c_{i}\rangle = \frac{H_0 f}{(2 \pi)^{3}}
  \int \frac{\mathbfit{k}\cdot \hat{\mathbfit{r}}}{k^{2}} P(k) e^{-i
    \mathbfit{k}\cdot\mathbfit{r}} \dd \mathbfit{k}
\end{equation}
It is also possible to directly
sample the velocity field using :
\begin{equation}
  \label{eq:10}
  v_{\alpha}^{CR} = v_{\alpha}^{RR} + \langle v_{\alpha} c_{i}\rangle \langle c_{i}c_{j}\rangle ^{-1}
(c_{i}-c^{RR}_{i})
\end{equation}
where the index $\alpha$ corresponds to the cartesian component. The
correlation $\langle v_{\alpha} c_{i}\rangle$ is given by
\begin{equation}
  \label{eq:8}
  \langle v_{\alpha} c_{i}\rangle = \sum_{\beta}\Psi_{\alpha,\beta} r_{\beta}
\end{equation}

In this
paper, we sample both the velocity and density field using the same
random realization. We thus don't need the periodic
boundary conditions implied by the use of fast Fourier transform to
evaluate Eq.~\ref{eq:vdef}.

\section{Test on mock}
\label{sec:mock}

\begin{figure}
  \centering
   \includegraphics[width=1\linewidth]{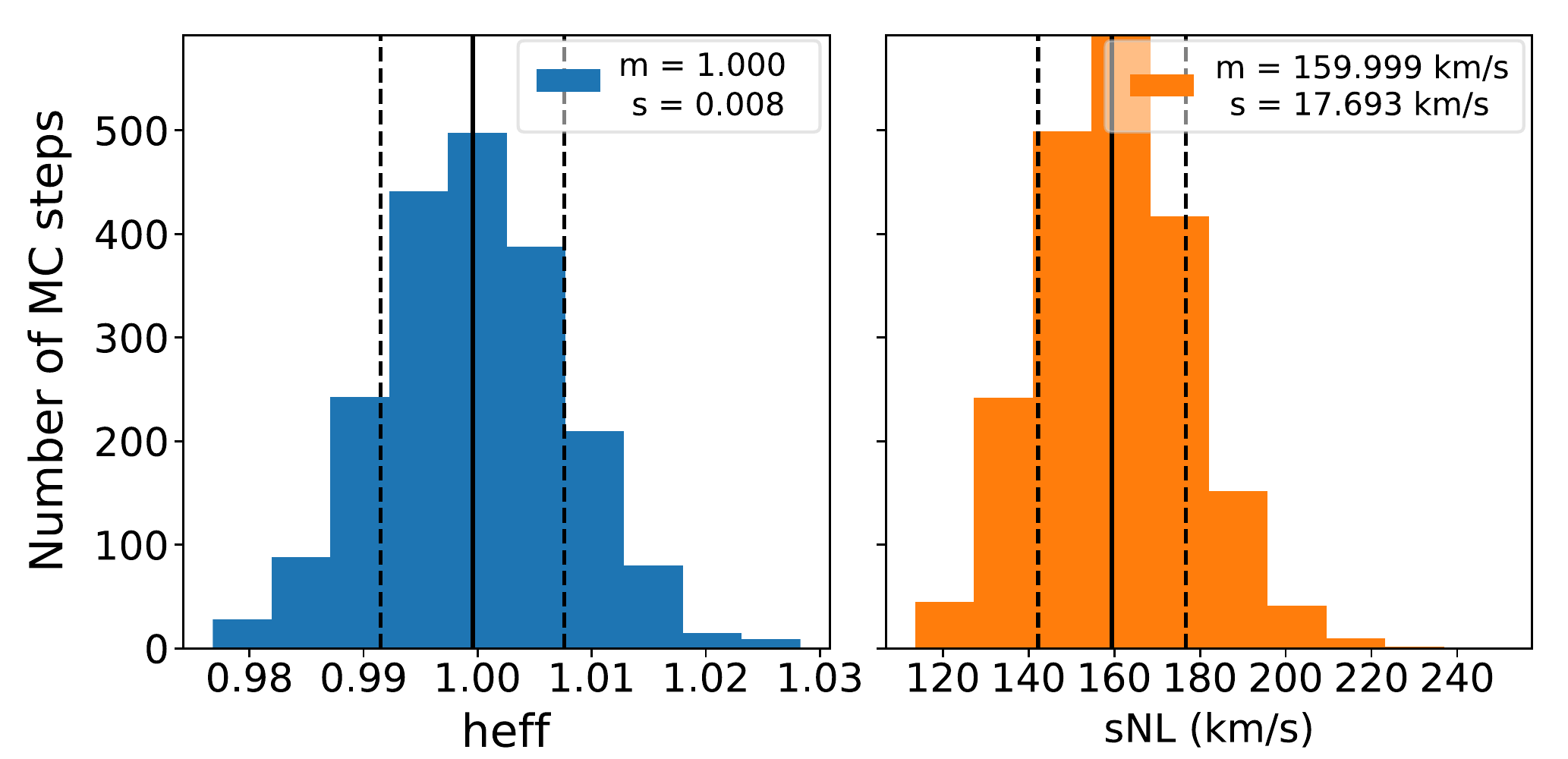}
  \caption{MCMC chain for the mock. The upper panel shows the
    $h_\text{eff}$ parameter while the bottom one shows the non-linear
    dispersion $\sigma_\text{NL}$.}
  \label{fig:mock_chain}
\end{figure}

\begin{figure}
  \centering
   \includegraphics[width=1\linewidth]{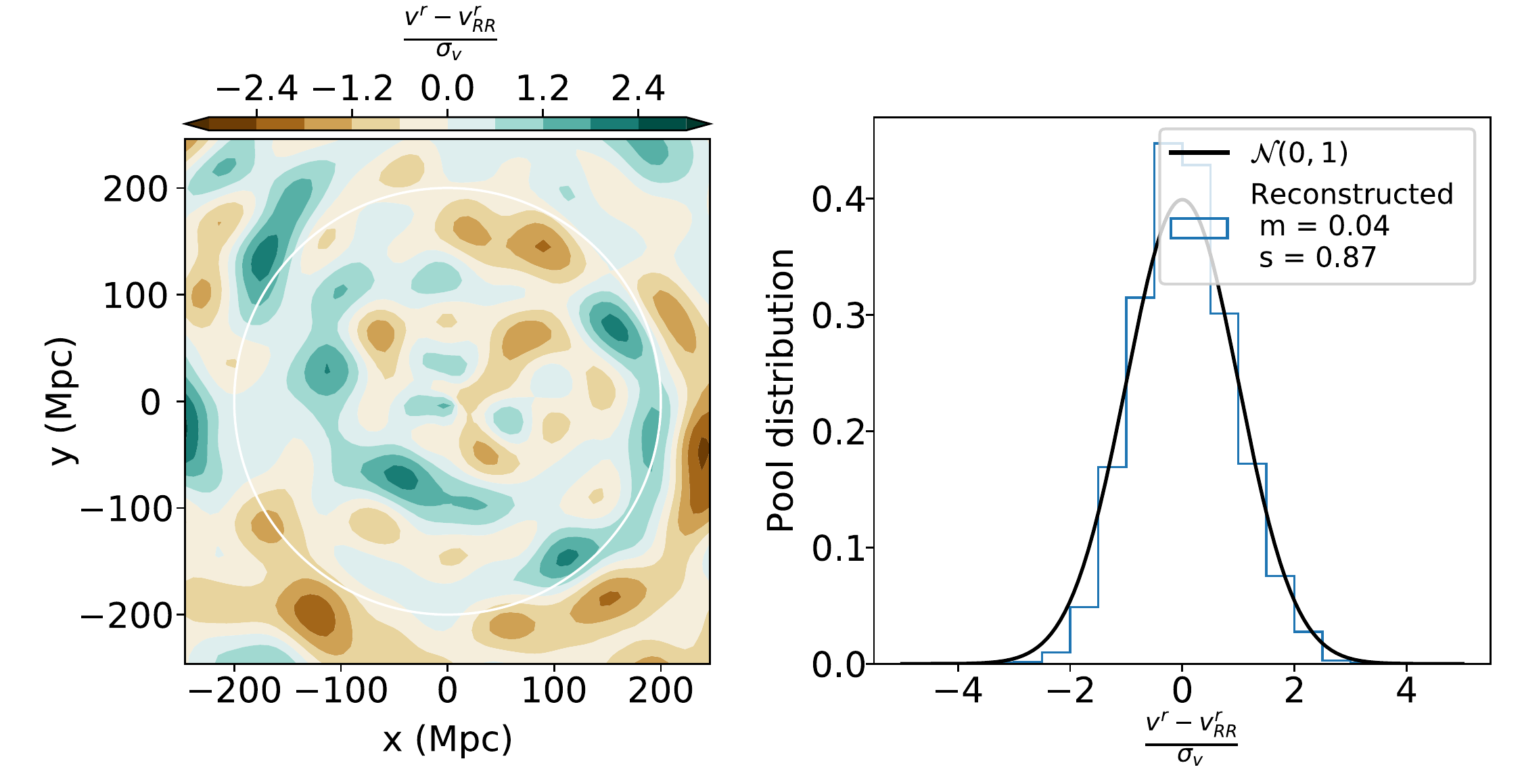}
  \caption{Comparison between the original radial peculiar velocities and
    the reconstructed ones. The left panel shows the residuals
    $\frac{v^{r}-v^{r}_{\text{RR}}}{\sigma_{v^{r}}}$ to the
    original field
    in the z=0 slice. The right panel shows the histogram of this
    quantity over the whole box. The black solid line is the unit Gaussian.}
  \label{fig:mock_comp}
\end{figure}

\begin{figure}
  \centering
   \includegraphics[width=1\linewidth]{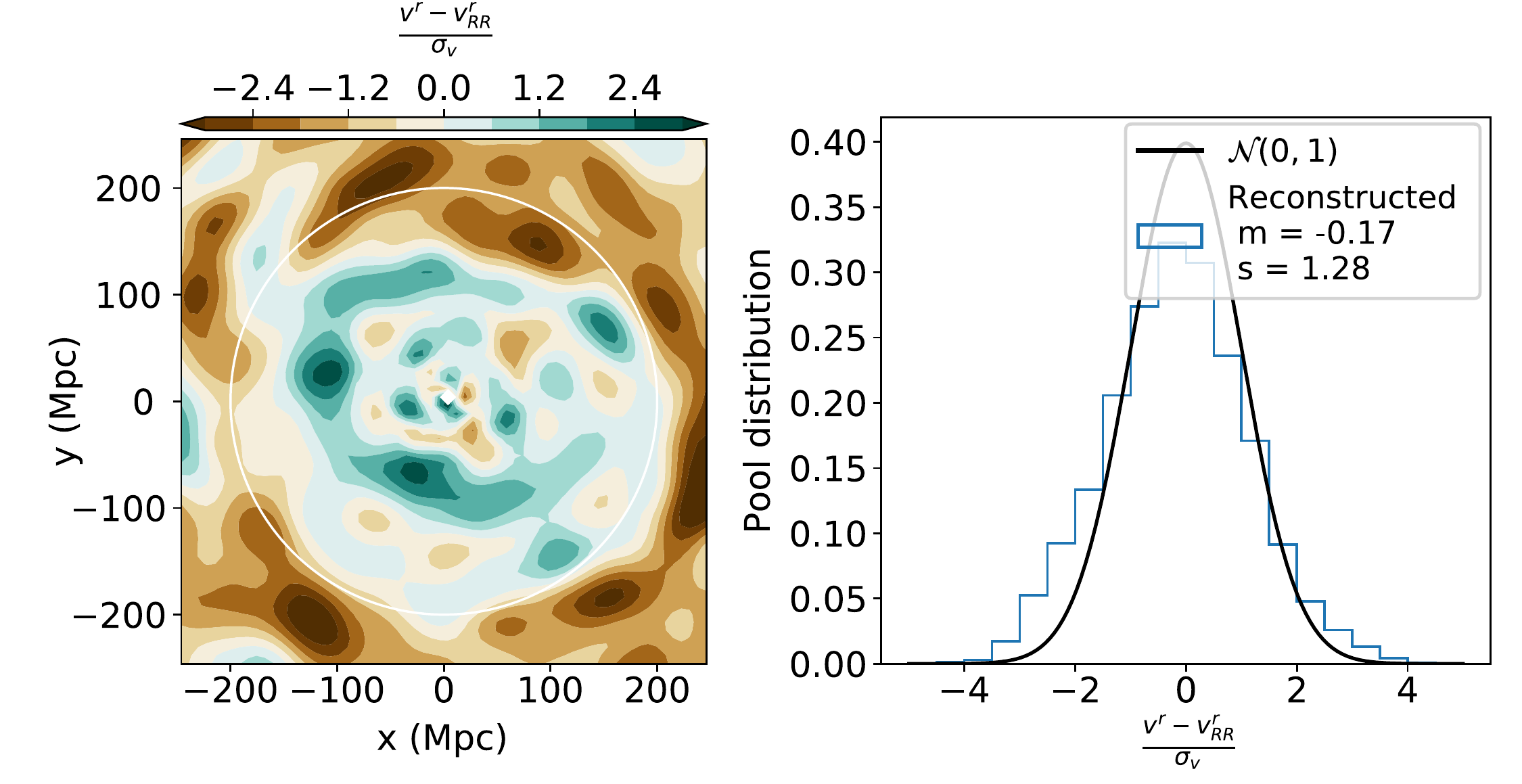}
  \caption{Same as Figure~\ref{fig:mock_comp} for the reconstruction using
    WF/CR technique alone.}
  \label{fig:WFCR_comp}
\end{figure}

We test the implementation of the algorithm on a mock catalog of 4000 tracers generated as follow :
\begin{itemize}
\item The angular positions are drawn from an uniform distribution,
\item The distances are drawn from a truncated normal distribution
  within $d \in [0,200]$ Mpc,
\item We mimic the observations by computing the corresponding distance
  moduli and scatter them following a normal distribution of standard
  deviation $\sigma_{\mu} = 0.2$,
\item We input a shift in the distance moduli scale of $h_{\text{eff}} =1.0$.
\item From a random realization generated from a power spectrum
  truncated at 0.1 Mpc (to avoid non-linearities), we draw the
  peculiar velocities. We add a non-linear dispersion with
  $\sigma_{\text{NL}}=150$ km/s,
\item From the original distances and peculiar velocities, we compute
  the measured redshifts and add a Gaussian dispersion of
  $\sigma_{cz}=50$ km/s,
\end{itemize}
From the simulated distance moduli and redshifts, we reconstruct the
velocity field following the procedure described in
Section~\ref{sec:sampling}. We compute it on a grid of size $64^{3}$ and
box of $500$ Mpc width. The Figure~\ref{fig:mock_chain} shows the
resulting distribution for the two parameters $h_{\text{eff}}$ and
$\sigma_{\text{NL}}$, which both agree with their fiducial values.
 The comparison between the
original Gaussian random field and the reconstructed radial velocity
field is shown on Figure~\ref{fig:mock_comp}. The left panel shows the
pull distribution of the reconstructed radial velocity field in the
$Z=0$ slice. On the left plot, we show the histogram of these values
within the white circle corresponding to the data limit. Overall the
distribution is close to a unit Gaussian, showing the unbiased aspect of
the reconstruction. Also, on the left panel, there is no clear sign of
reconstructed structures over more than $3\sigma$. 
For comparison, the same plot is done using the WF/CR technique alone,
\emph{i.e} fixing the parameters $d$, $\sigma_{\text{NL}}$ and
$h_{\text{eff}}$, is shown on Figure~\ref{fig:WFCR_comp}. We can see the strong improvement from the
method. Homogeneous Malmquist bias affects the distances so that there
is excessive outflow in the center of the box and inflow outside, as it
is described in Section~\ref{sec:malmquist-bias-1}. 

\section{Changing the distance priors}
\label{sec:chang-dist-priors}
\begin{figure}
  \centering
   \includegraphics[width=1\linewidth]{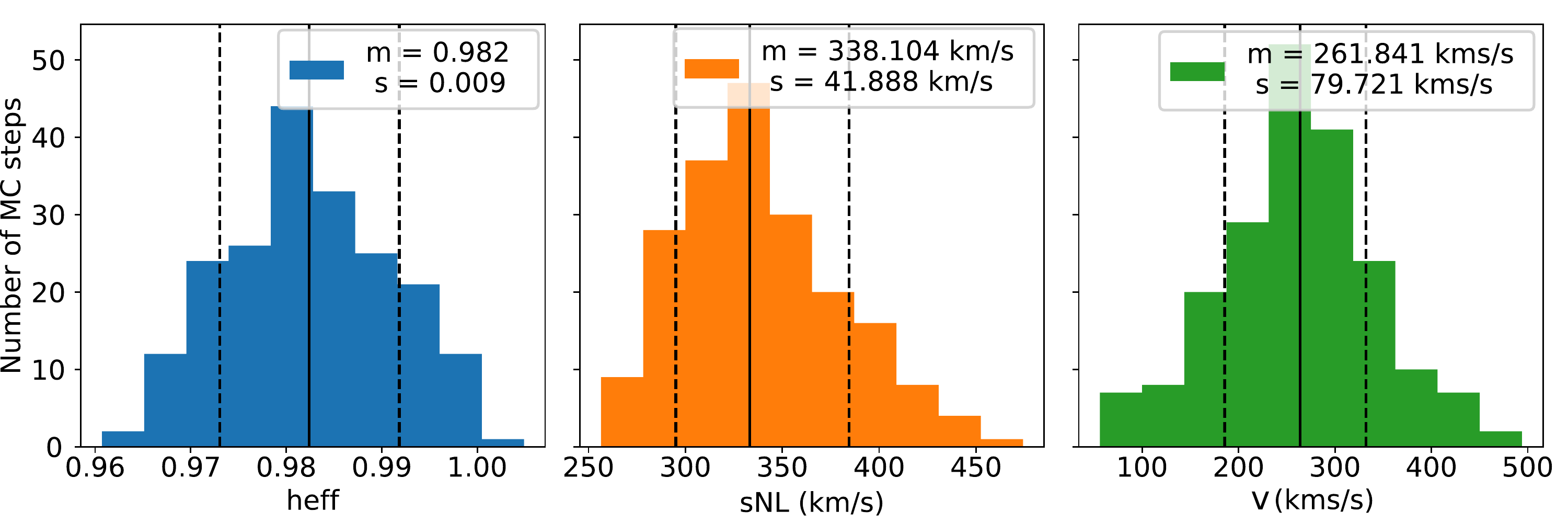}
  \caption{The posterior distribution for three parameters in the case of a
    reconstruction considering uniform priors en distances (except for
    6dF data, see text):  (left) 
    the effective reduced Hubble constant $h_{\text{eff}}$; (mid) the non linear
    dispersion $\sigma_{\text{NL}}$; (right) the reconstructed radial velocity
    at Virgo, of coordinates $(SGX,SGY,SGZ) =  (-3.6, 15.6, -0.7)$
    Mpc.}
  \label{fig:hemp_chain}
\end{figure}

\begin{figure*}
  \centering
     \includegraphics[width=1\linewidth]{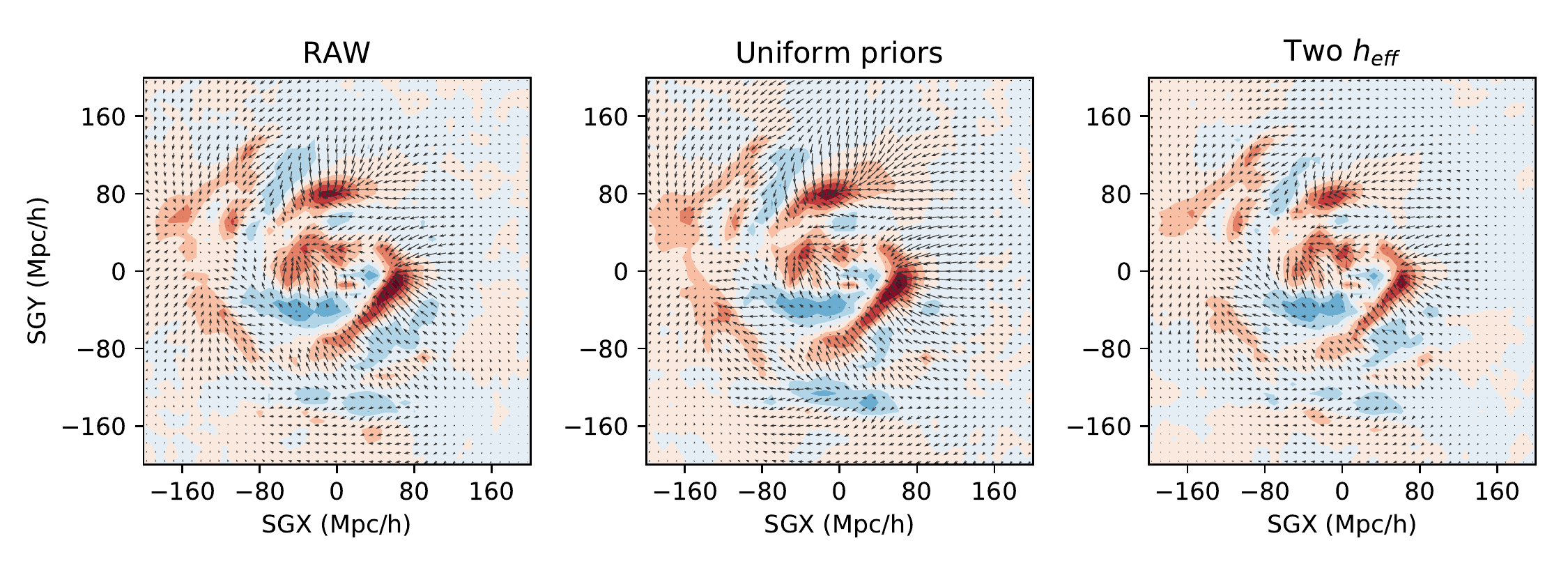}

  \caption{Same as Figure~\ref{fig:delta_1H} in the case of three
    different 
    reconstructions from the half CF3 catalog (see text). Left: Nominal
    reconstruction; mid: reconstruction with uniform priors except for
  6dF subsample; right: reconstruction with an independent zero-point
  shift for 6dF subsample.}
  \label{fig:comparison}
\end{figure*}

We test the effect of changing the assumed priors on distances on a
subsample of the CF3 catalog. The test catalog is built by randomly
taking half of the galaxies in each subsample defined in
Section~\ref{sec:selec}. The reconstruction is done on a grid of size
$64^{3}$. The reconstruction was carried in the case of uniform priors
imposed on all distances except the 6dF one. We treat 6dF as separate
because the sharp cutoff on their redshift makes the analysis unrealistic
with uniform prior. On 6dF, we fit an empirical
prior defined by Eq.~\ref{eq:emp}. The calculation was carried on 800
MCMC steps and the first 300 are considered as the warmup phase. The
Figure~\ref{fig:hemp_chain} shows the resulting histograms for the
$h_{\text{eff}}$ and $\sigma_{\text{NL}}$ parameters. One can see that
the effective Hubble constant has been minored compared to the
reconstruction shown in Figure~\ref{fig:hist_1H} by more than one sigma. This shows the strong
dependance of any Hubble constant determination with this method on the
prior distance distributions. Figure~\ref{fig:comparison} shows the
comparison between the reconstructed fields in the case of a nominal
reconstruction (empirical priors and unique shift in the zero-point)
(left) and the case of uniform priors (middle). The differences are
small, but one can notice the inflow on the $SGX>0$ Mpc/$h_{75}$ part (data
which is mainly not covered by 6dF data) has been increased. Because we
imposed no priors on the data, the distances are more likely to be
overestimated far away, and consequently the radial velocities
underestimated, biasing the field towards negative values. This
results shows that the prior distribution has an impact in our
Bayesian analysis and underlines the fact that
selection effects have to be properly modeled to extract
cosmological parameters from the velocity field reconstruction.


\section{Considering 6dF as a separate subsample}
\label{sec:2H}
\begin{figure}
  \centering
   \includegraphics[width=1\linewidth]{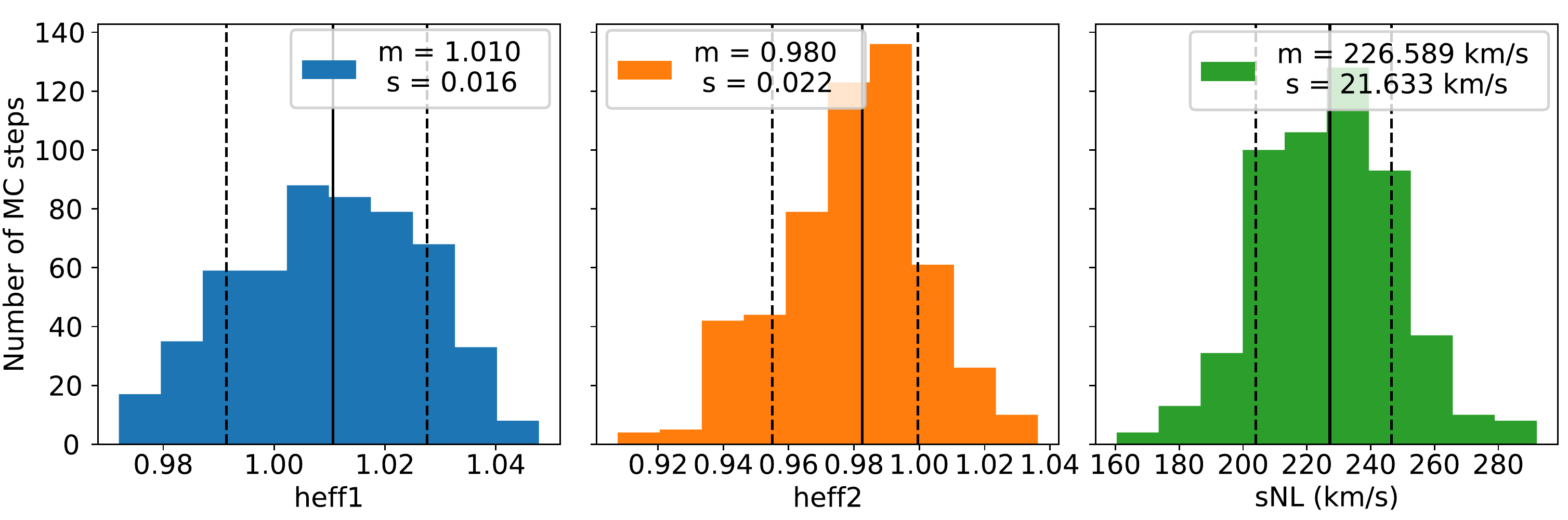}
  \caption{The posterior distribution for three parameters in the case of a
    reconstruction with two different effective Hubble constants (see text): (left) 
    the effective reduced Hubble constant $h_{\text{eff}}$ for 6dF data;
    (mid) $h_{\text{eff}}$  for the other galaxies; (right) the non linear
    dispersion $\sigma_{\text{NL}}$.}
  \label{fig:hist_2H}
\end{figure}

We try to fit two different reduced effective Hubble constants $h_{\text{eff}}$, one for 6dF
data and the other for the rest of the data. The distance priors are
kept to empirical prior functions defined by  Eq.~\ref{eq:emp}. The
resulting histograms are shown on Figure~\ref{fig:hist_2H}. We see that
the values recovered for the two $h_{\text{eff}}$ parameters are
different while still compatible. We can see
that the non-linear dispersion has decreased compared to
the main and uniform priors reconstructions ($\sim 225$ km/s compared to
$\sim 270$ km/s). This suggests a better fit of the underlying velocity
field. However, it could also be due to the artificial reduction of a
real flow between 6dF covered region and the rest, which is why the main
result of this article assumes a unique zero-point shift. Again, Figure~\ref{fig:comparison} shows the
comparison between the reconstructed fields in the case of a fiducial
reconstruction (empirical priors and unique shift in the zero-point)
(left) and the case of two different $h_{\text{eff}}$ (right). The
differences are more pronounced: The inflows in the regions $SGX>100$
Mpc/$h_{75}$ and $SGX<200$ Mpc/$h_{75}$ have been reduced. Also, we can see from the overdensity field that
the overall inflows on structures have changed between the left and right panels,
increasing in the case of a unique zero-point. We stress that the
difference of flows can be artificial and the conservative approach
taken in the main result of this paper (empirical priors and unique
zero-point shift) should be preferred. However, this test shows that
small shifts in distance indicators calibrations change the mean inflow
on structures, and could have an impact on the determination of
cosmological parameters, such as the growth rate
$f \sigma_{8}$ or the Hubble constant $H_{0}$. This impact is however
beyond the work presented in this paper. Extending the model to
the determination of cosmological papers from peculiar velocities is to be considered as a possible
development of the current methodology.

\bsp    
\label{lastpage}
\end{document}